\documentclass[journal,article,submit,moreauthors,pdftex,10pt,a4paper]{mdpi} 
\firstpage{1} 
\articlenumber{x}
\doinum{10.3390/------}
\pubvolume{xx}
\pubyear{2018}
\copyrightyear{2018}
\history{Received: date; Accepted: date; Published: date}
\preto{\abstractkeywords}{\nolinenumbers}

\Title{Two novel approaches to the hadron-quark mixed phase in compact stars}

\Author{Vahagn Abgaryan$^{1}$\orcidA{}, David Alvarez-Castillo$^{2}$\orcidB{}*, Alexander Ayriyan$^{1}$\orcidC{}**,\\ David Blaschke$^{2,3,4}$\orcidD{}, Hovik Grigorian$^{1,5}$\orcidE{}}

\AuthorNames{Vahagn Abgaryan, David Alvarez-Castillo, Alexander Ayriyan, David Blaschke, Hovik Grigorian}

\address{%
$^{1}$ \quad Laboratory for Information Technologies,
	Joint Institute for Nuclear Research,
	Joliot-Curie street 6,
	141980 Dubna, Russia\\
$^{2}$ \quad Bogoliubov Laboratory for Theoretical Physics,
	Joint Institute for Nuclear Research,
	Joliot-Curie street 6,
	141980 Dubna, Russia\\
$^{3}$ \quad Institute of Theoretical Physics, 
	University of Wroclaw, 
	Max Born place 9, 
	50-204 Wroclaw, Poland\\
$^{4}$ \quad National Research Nuclear University (MEPhI),
	Kashirskoe Shosse 31,
	115409 Moscow, Russia\\
$^{5}$ \quad Department of Physics, 
	Yerevan State University, 
 	Alek Manukyan Str. 1, 
 	0025 Yerevan, Armenia}

\corres{Correspondence: $^{*}$~alvarez@theor.jinr.ru; $^{*}$~ayriyan@jinr.ru}

\abstract{
First-order phase transitions, like the liquid-gas transition, proceed via formation of structures such as bubbles and droplets. In strongly interacting compact star matter, at the crust-core transition, but also at the hadron-quark transition in the core, these structures form different shapes dubbed "pasta phases". 
We describe	two methods to obtain one-parameter families of hybrid equations of state (EoS) which mimic the thermodynamic behavior of pasta phases in between a low-density hadron and a high-density quark matter phase, thus generalizing the Maxwell construction.
The first method replaces the behavior of pressure vs. chemical potential in a finite region around the critical 
pressure of the Maxwell construction by a polynomial interpolation.
The second method uses extrapolations of the hadronic and quark matter EoS beyond the Maxwell point to define a mixing of both with weight functions bounded by finite limits around the Maxwell point.
We apply both methods to the case of a hybrid EoS with a strong first order transition that entails the formation of a third family of compact stars and the corresponding mass twin phenomenon.
We investigate for both models the robustness of this phenomenon against variation of the single parameter, the pressure increment at the critical chemical potential which quantifies the deviation from the Maxwell construction.
We also show sets of results for other compact star observables than mass and radius, namely the moment of inertia and the baryon mass.	
}


\usepackage{lipsum}
\usepackage{graphicx,pstricks,epsfig,rotating,amsmath,amssymb}

\newcommand{\bea}{\begin{eqnarray}}
\newcommand{\eea}{\end{eqnarray}}
\newcommand{\apgt} {\ {\raise-.5ex\hbox{$\buildrel>\over\sim$}}\ }

\begin{document}


\section{Introduction}

The understanding of the properties of dense matter in compact star interiors is a subject of current research. Recently, great progress in this direction has been achieved by the detection of the gravitational radiation 
which emerged from the inspiral phase of two coalescing compact stars, an event named GW170817~\cite{TheLIGOScientific:2017qsa}. Since it was observed also in all other bands of the electromagnetic spectrum, it marked the birth of multi-messenger astronomy.
Among the various obtained results, GW170817 has shed light on the properties of the equation of state (EoS) of compact star matter, namely on its stiffness, since through the constraints on the tidal deformability parameter $\Lambda$ \cite{Hinderer:2009ca} from the LIGO-Virgo Collaboration (LVC) results one could estimate the maximum radius of a 1.4~M$_\odot$ compact star to $R_{1.4,{\rm max}}=13.6$ km \cite{Annala:2017llu} and maximum mass of nonrotating compacts stars $M_{\rm TOV, max}=2.16$ M$_\odot$ \cite{Rezzolla:2017aly}.
Of great scientific interest is the phenomenon of a phase transition from hadronic matter to a deconfined quark phase in hybrid compact stars. Those stars are comprised of a deconfined quark matter core surrounded by a hadronic mantle. The nature of the deconfinement transition is a matter of debate. 
Whether it exhibits a jump in the thermodynamic variables or represents a crossover\footnote{Here the term "crossover" is used generically for a transition that does not proceed like in a Maxwell construction at a strictly constant pressure with a jump in (energy) density, but rather by a varying pressure in the transition region. It can thus be a generic crossover transition like in ferromagnetic systens under external magnetic field, but also a first order transition for several globally conserved charges which proceeds via formation of structures of different shapes (pasta phases).} is a question that is addressed to both, laboratory experiments as well as compact star observations.
The possible mixed phase in a neutron star presents geometrical structures of different shapes, in the literature denoted as "pasta phases". 
The adequate description of the physics involved is a complicated problem where the geometrical properties of the structures as well  as transitions between them must be taken into account (different methodologies can be found in~\cite{Voskresensky:2002hu,Maruyama:2007ss,Watanabe:2003xu,Horowitz:2005zb,Horowitz:2014xca,Newton:2009zz,Yasutake:2014oxa}). 
In the case of the hadron-quark interface, the procedure is well explained in~\cite{Maruyama:2007ey} (see also \cite{Yasutake:2014oxa} for a recent work): one models several geometrical structures and finds the energetically most favorable ones in different density regions inside compact stars. 

In this work we take a different route and introduce two types of phenomenological interpolations 
which aim at mimicking the effect of those geometrical structures on the thermodynmical behavior and simultaneously explore the whole range of densities in a unified way. 
A first realization the idea to describe the transition from the hadronic to the quark matter phase of matter in neutron stars by an interpolation in order to model a crossover-like behaviour was carried out in \cite{Masuda:2012ed} and followed up in Refs.~\cite{Alvarez-Castillo:2014dva,Alvarez-Castillo:2017xvu}, where
the jump of the EoS $\varepsilon(P)$ was replaced by a smooth behavior using as an ansatz a tangens hyperbolicus function.

\begin{figure*}[!htb]
	\begin{center}
		\includegraphics[width=0.55\textwidth]{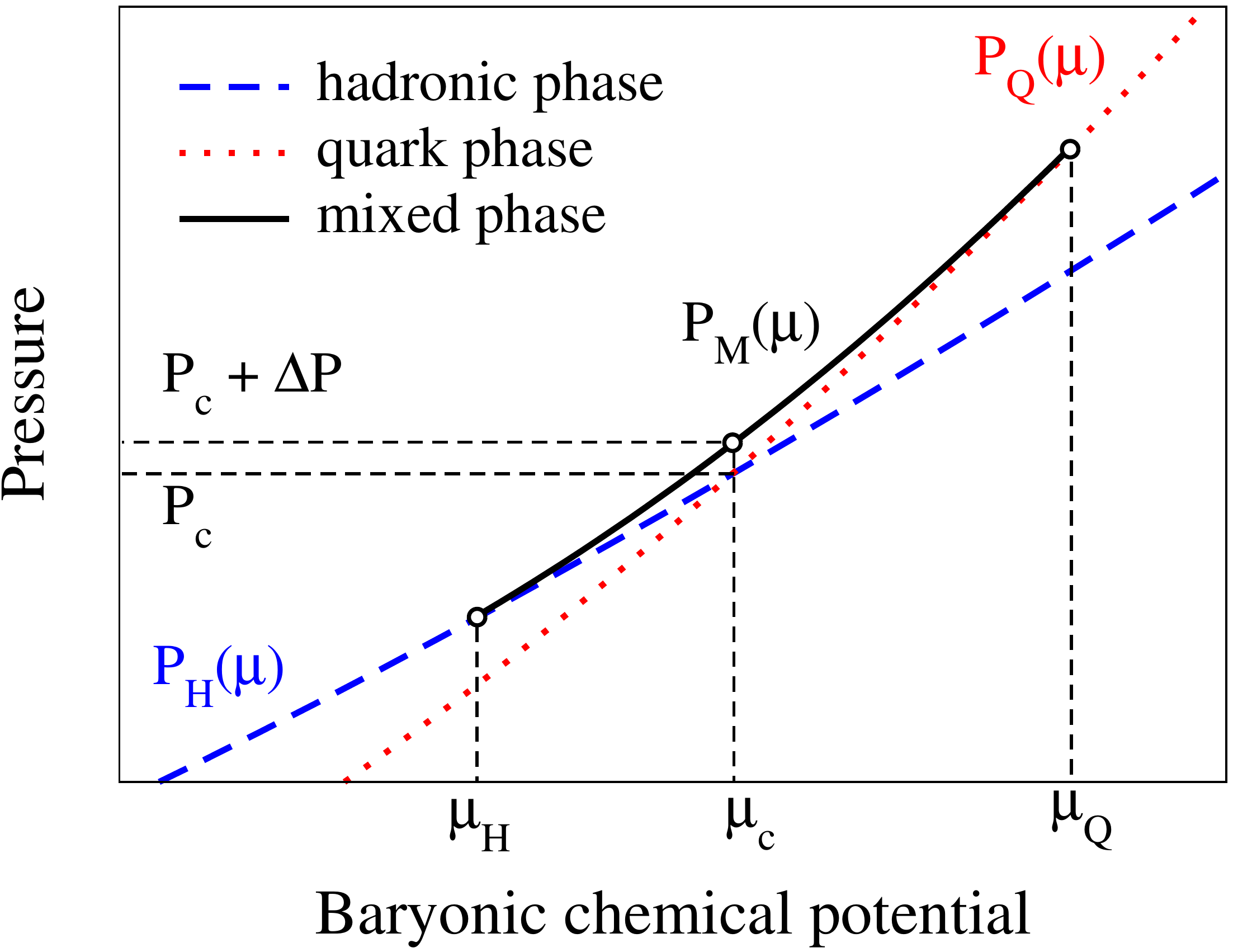} 
	\end{center}
	\vspace {-5mm}
	\caption{\label{idea}
		Schematic representation of the interpolation function $P_M(\mu)$ obtained from the mixed phase constructions discussed in this work. For both interpolation methods discussed in the text it has to go though three points: $P_H(\mu_{H})$, $P_{c}+\Delta P$ and  $P_Q(\mu_{Q})$.} 
\end{figure*}

A systematic and thermodynamically consistent formulation was recently given in 
\cite{Ayriyan:2017nby,Ayriyan:2017tvl}
where a parabolic interpolation function was introduced to {\it replace} the behavior of the hybrid EoS for a Maxwell transition. 
We shall denote this procedure as replacement interpolation method (RIM).
The resulting hybrid EoS  was then used to study the effect of the mixed phase on the properties of compact stars. 
A second realization of this concept has been worked out recently in \cite{Alvarez-Castillo:2018rrv}, where instead of replacing the hadronic and quark matter branches of the hybrid EoS in the limits $\mu_H<\mu<\mu_Q$ (see  Fig.~\ref{idea}) a {\it mixing} of these branches is defined using switch functions and a bell-shaped function for the pressure increment with an amplitude $\Delta P=\Delta_P\, P_c$, where $P_c=P(\mu_c)$ is the critical pressure of the Maxwell construction. 
This procedure is denoted as mixing interpolation method (MIM) in \cite{Alvarez-Castillo:2018rrv}. 
The free parameter $\Delta_{P}$ occurs in both methods with an equivalent influence on the behaviour of the EoS in the mixed phase region, in particular on its extension, see Fig.~\ref{idea}. We would like to note that in both methods a negative value of  $\Delta_{P}$ would signal that a Maxwell construction using both input EoS $P_H(\mu)$ for hadronic matter and $P_Q(\mu)$ for quark matter would not make sense because it would describe a transition from quark matter at low densities 
(where $P_Q(\mu)$ is not trustworthy) to hadronic matter at high densities (where $P_H(\mu)$ is not trustworthy). For a discussion of this situation, see Ref.~\cite{Kojo:2014rca}.

In this work we present a comparative study of the RIM and MIM approaches to construct mixed phases of the quark-hadron phase transition which mimic the thermodynamic behaviour of pasta phases. 
We discuss the similarities and differences of these two approaches and apply them to obtain a hybrid EoS under neutron star constraints for which we discuss the resulting hybrid star sequences and their properties. 
While the first approach (RIM) is rather intuitive and simple to realise as its properties just depend on the order of interpolating polynomial, the second approach (MIM) is based on a procedure of "mixing" the EoS of the two phases in the coexistence region and reminds in its properties on the physics of substitutional compounds as in the crust of compact stars, resulting in an intermediate stiffening effect.

The paper is structured as follows. 
In section \ref{sec:hEoS} we start with the reference EOS for the present study, for which a four-polytrope ansatz is employed which features a hadronic phase (first polytrope), a constant pressure polytrope resembling a strong first order phase transition as described by a Maxwell construction (second polytrope) and two polytropes for quark matter phases at high densities. 
Next, in section \ref{sec:mixed} we introduce the RIM and MIM approaches to construct mixed phases when two reference EoS for the low-density (hadronic) and high-density (quark matter) phases are given. We discuss the speed of sound $c_s$ as the key characterizing property of the family of obtained hybrid EoS. 
Subsequently, in section \ref{sec:results} we discuss the similarities and differences between the hybrid star EoS of both approaches and show results for the macroscopic properties of compact stars. 
We motivate these results by the feasibility of detection by multi-messenger astronomy. 
Consequently, future detections of gravitational wave radiation emitted by of NS-NS or NS-BH mergers shall provide new constraints on both the star mass and radius. 
Moreover, the determination of the fate of the merger, whether it evolves via a prompt or delayed collapse into a black hole, can be used as an independent estimate on the mass and radius, as proposed in~\cite{Bauswein:2017vtn}. 
Up to now, tests for the current compact star models with the at present still single compact star merger event have been performed, e.g., in~\cite{Paschalidis:2017qmb,Ayriyan:2017nby,Alvarez-Castillo:2018rrv}.

\section{Hybrid star EoS with a third family and high-mass twins}
\label{sec:hEoS}

Compact stars are traditionally divided into white dwarf (first family) and neutron star (second family) branches. 
Hybrid stars whose equation of state undergoes a sufficiently strong first order phase transition (large jump in energy density $\Delta \varepsilon$) can populate a third family branch in the mass-radius diagram, separated from the second one by a sequence of unstable configurations.  
As a consequence, there appear so called mass twin configurations: the second and third family solutions overlap within a certain range of masses while the radii of any two stars with the same mass (mass twins) are very different. 
If the mass-twin phenomenon occurs at high masses $\sim 2$~M$_\odot$ then one speaks of high-mass twin (HMT) 
stars~\cite{Benic:2014jia}.
Depending on the critical pressure of the phase transition, the mass-twin phenomenon can occur also at lower masses such as the typical binary radio pulsar mass of $\sim 1.35$~M$_\odot$, see \cite{Paschalidis:2017qmb,Ayriyan:2017nby,Alvarez-Castillo:2018rrv}, so that the corresponding twin star configuration become of relevance for the interpretation of GW170817. 
In the latter case, a mass ratio $q=m_1/m_2=1$ of the merger would not entail that the merging stars have the same radii and internal structure!
Would the mass-twin phenomenon (at whatever mass) be observed, this would entail that the QCD phase diagram has to possess at 
least one critical endpoint since for the study of the cold region of the QCD phase diagram the existence of a first order phase transition between hadron to quark matter had to be concluded. 
Since the high temperature region of the QCD diagram is kown to feature a crossover transition, compact stars can serve as a probe of the existence of a critical end point~\cite{Alvarez-Castillo:2015xfa} and provide insight into the properties of matter in heavy ion collision conditions~\cite{Alvarez-Castillo:2016wqj}. 

In order to study the effects of pasta phases at the hadron-quark matter interface in hybrid star interiors, we consider a piecewise polytropic EoS as previously used in various works~\cite{Read:2008iy,Hebeler:2013nza,Raithel:2016bux,Alvarez-Castillo:2017qki,Annala:2017llu}. 
The polytropic representation used in the present work consists of four segments of matter at
densities higher than saturation density $n_0=0.15$ fm$^{-3}$ ($n_0 \ll n_1<n<n_5$).

\bea
\label{polytrope}
P(n) =  \kappa_i  (n/n_0)^{\Gamma_i}, \  n_i < n < n_{i+1}, \ i=1 \dots 4,
\eea 

Each density region is labeled by $i=1\dots 4$ with prefactor $\kappa_i$ and  polytropic index  $\Gamma_i$. 
HMT stars require a rather stiff nucleonic EoS which here is represented by the first polytrope. 
The hadron-quark matter first-order phase transition is described by the second polytrope with constant pressure $P_{tr}=\kappa_2$ and vanishing polytropic index ($\Gamma_2=0$). 
At higher densities the polytropes 3 and 4 represent a rather stiff quark matter EoS. 
The parameters for this HMT realisation are given in table~\ref{param-123}. 
\begin{table}[!htb]
	\centering
	\caption{
		Parameters for the four-polytrope EoS of Ref.~\cite{Alvarez-Castillo:2017qki}, called "ACB4" in Ref.~\cite{Paschalidis:2017qmb}.  
		The corresponding description is presented in Eq.~\ref{polytrope} of the main text. 
		The last column displays the maximum masses $M_{\rm max}$ on the hadronic
		(hybrid) branch corresponding to region $i=1$ ($i=4$). In addition, the minimal
		mass $M_{\rm min}$ in region $i=3$ of the hybrid branch is displayed in that column.}
	\label{param-123}
	\begin{tabular}{c|c|cccc|c}
		\hline \hline
		&&$\Gamma_i$&$\kappa_i$&$n_i$ &$m_{0,i}$&$M_{\rm max/min}$\\		
		ACB&i&&[MeV/fm$^3$]&[1/fm$^3$] &$[MeV]$&$[M_\odot$]\\
		\hline
		4&1&4.921&2.1680&0.1650&939.56 & 2.01  \\
		&2&0.0&63.178&0.3174&939.56 & -- \\
		&3&4.000&0.5075&0.5344&1031.2 & 1.96  \\
		&4&2.800&3.2401&0.7500&958.55 & 2.11  \\
		\hline \hline
	\end{tabular}
	\label{tab:1}
\end{table}
For the present applications to thermodynamically consistent interpolating constructions we need to convert the EoS (\ref{polytrope}) to the form \cite{Alvarez-Castillo:2017qki}
\begin{eqnarray}
\label{P-mu}
P(\mu)=\kappa_i\left[(\mu - m_{0,i})\frac{\Gamma_i -1}{\kappa_i\Gamma_i} \right]^{\Gamma_i/(\Gamma_i-1)} ~,
\end{eqnarray}
valid for the respective regions (phases) $i=1\dots 4$, where for the constant pressure region $i=2$
this formula collapses to $P(\mu=\mu_{c})=P_{c}=\kappa_2$ because of $\Gamma_2=0$.
For applying the MIM below, it will be important that the pressure of the hadronic phase ($i=1$) 
valid for $\mu<\mu_{c}$ can be extrapolated to the neighboring quark matter phase ($i=3$) where 
$\mu>\mu_{c}$ and vice-versa.

HMT star EoS fulfill the Seidov conditions over quantity values at the phase transition~\cite{Seidov:1971}
\bea
\label{seidov}
\frac{\Delta\varepsilon}{\varepsilon_{c}} \ge \frac{1}{2} 
+ \frac{3}{2} \frac{P_{c} }{\varepsilon_{c}} 
\eea
for the third family of compact stars to exist. 
The choice of parameters for this EoS corresponds to a sufficiently stiff high-density region in order to prevent gravitational collapse while at the same time not violating the causality condition for the speed of sound $c_s<c$. See~\cite{Alvarez-Castillo:2017qki} for details.

\section{Mixed phase constructions}
\label{sec:mixed}

In this section we present the details of the interpolation descriptions for the mixed phase between the hadronic and quark matter phases. 
For this purpose we consider the chemical potential dependent pressures of both the hadronic ($i=1$) and the neighboring  quark matter ($i=3$) phases: $P_{H}(\mu)$, $P_{Q}(\mu)$, respectively.  
As mentioned above, our polytropic HMTs EoS features a first order phase transition implemented in the form of a Maxwell construction at  a critical chemical potential value $\mu_c$ where pressures for both phases are equal:
\begin{equation}
P_{Q}\left(\mu_{c}\right)=P_{H}\left(\mu_{c}\right) = P_c,
\end{equation}
thus both phases are in thermodynamic equilibrium. 


\subsection{The replacement interpolation method (RIM)}
\label{ssec:RIM}

In this mixed phase approach the relevant regions of both, the hadronic and quark matter EoS around the Maxwell critical point ($\mu_{c},P_{c}$) are replaced by a polynomial function defined as
%
\begin{equation}
\label{mph_general}
P_{M}\left(\mu\right) = \sum_{q=1}^{N} \alpha_{q} \left(\mu - \mu_{c}\right)^{q} + \left(1 + \Delta_{P}\right)P_{c}
\end{equation}
where $\Delta_{P}$ is a free parameter representing additional pressure of the mixed phase at $\mu_{c}$.
Generally, the ansatz (\ref{mph_general}) for the mixed phase pressure is an even order ($N = 2 k$, k=1,2, ...) 
polynomial and it smoothly matches the EoS at $\mu_{H}$ and $\mu_{Q}$ up to the $k$-th derivative of the pressure,
\begin{eqnarray}
\label{press}
    P_{H}\left(\mu_H\right) &=& P_{M}\left(\mu_H\right)~,~~
    P_{Q}\left(\mu_Q\right) = P_{M}\left(\mu_Q\right)\\
    \frac{\partial^q}{\partial \mu^q} P_{H}\left(\mu_H\right) &=& \frac{\partial^q}{\partial \mu^q} P_{M}\left(\mu_H\right)~,~~
	\frac{\partial^q}{\partial \mu^q} P_{Q}\left(\mu_Q\right) = \displaystyle \frac{\partial^q}{\partial \mu^q} P_{M}\left(\mu_Q\right)~,~~q=1,2,\dots,k~,
\label{press-deriv}
\end{eqnarray}
where $N+2$ parameter values ($\mu_{H}$, $\mu_{Q}$ and $\alpha_{q}$, for $q=1,\dots,N$) can be found by solving the above system of equations, leaving one parameter ($\Delta P$) as a free parameter of this method.

The simplest case of the RIM is the parabolic model for $N=2$ which has been first introduced in \cite{Ayriyan:2017tvl,Ayriyan:2017nby},
\begin{equation}
P_{M}\left(\mu\right) = \alpha_{2} \left(\mu - \mu_{c}\right)^{2} + \alpha_{1} \left(\mu - \mu_{c}\right) 
      + \left(1 + \Delta_{P}\right)P_{c}
\end{equation}
As usual, the parameters $\alpha_{1}$, $\alpha_{2}$, $\mu_{H}$ and $\mu_{Q}$ are found from the following system of equations involving quantities at the borders of the mixed phase,
\begin{eqnarray}
    P_{H}\left(\mu_H\right) &=& P_{M}\left(\mu_H\right)~,~~
    P_{Q}\left(\mu_Q\right) = P_{M}\left(\mu_Q\right)\\
    n_{H}\left(\mu_H\right) &=& n_{M}\left(\mu_H\right)~,~~
    n_{Q}\left(\mu_Q\right) = n_{M}\left(\mu_Q\right)~.
\end{eqnarray}
It is evident that the order of the interpolating function (\ref{mph_general}) will determine whether or not there are discontinuities for the derivatives of the function $P_{M}(\mu)$. 

\begin{figure*}[!thb]
	\begin{center}$
		\begin{array}{cc}
		\includegraphics[width=0.5\textwidth]{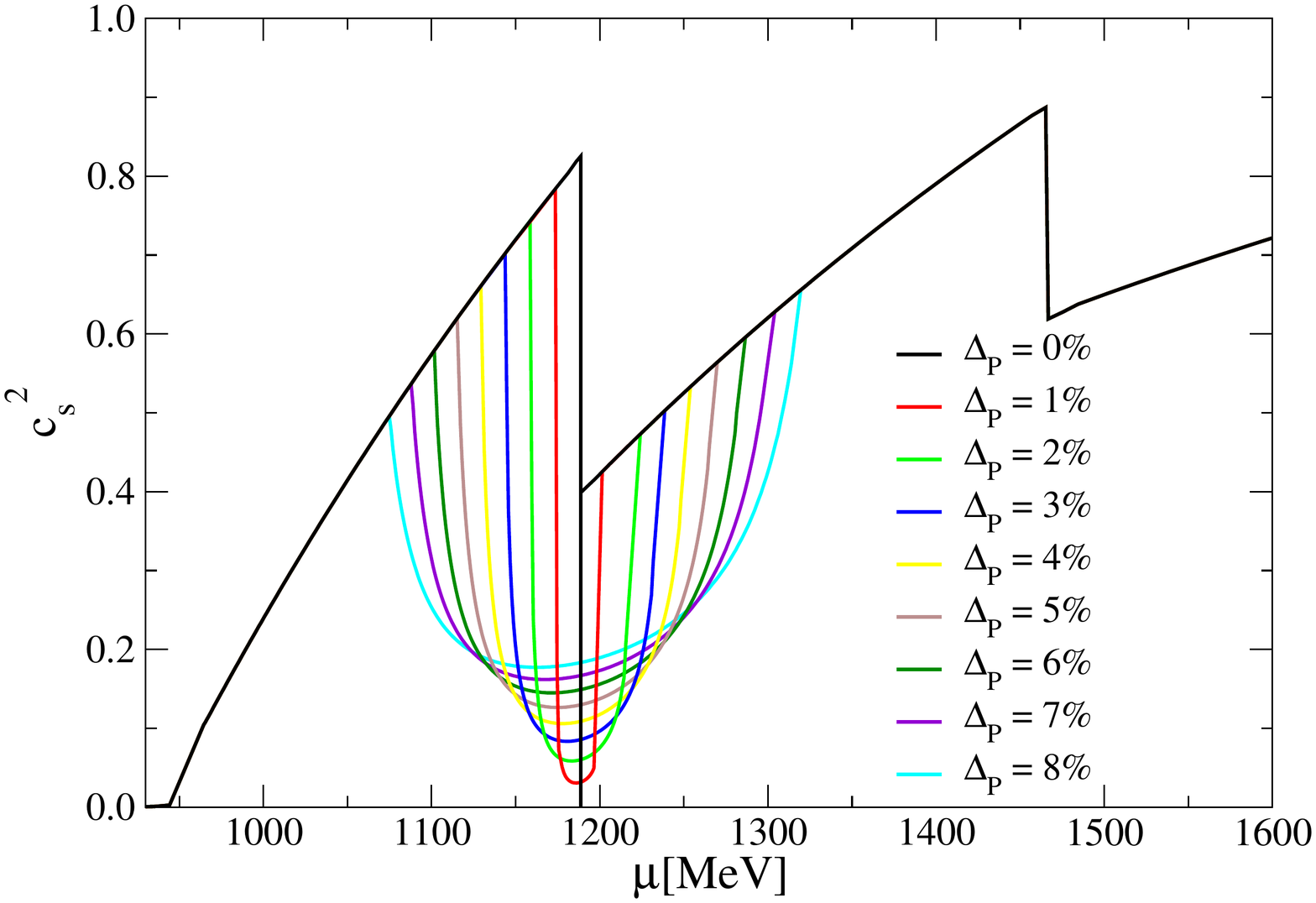} & \hspace{0cm}\includegraphics[width=0.5\textwidth]{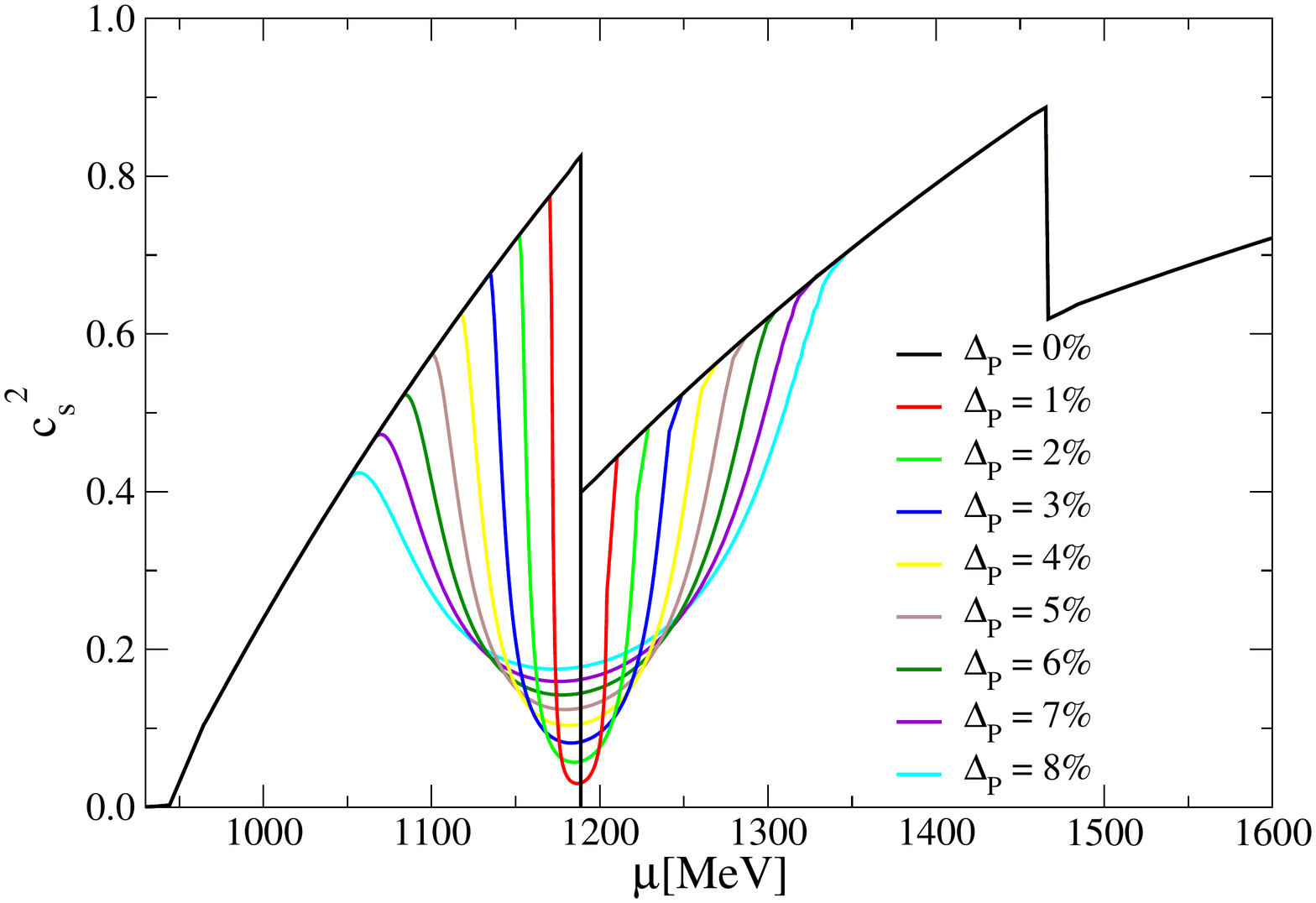}
		\end{array}$
	\end{center}
	\caption{The squared speed of sound as a function of the chemical potential for the MIM construction with $k=2$ (left panel) and $k=3$ (right panel). 
	} 
	\label{fig:g2} 
\end{figure*}

For instance, the square of the speed of sound, 
\begin{equation}
c_s^{2}=\frac{\partial P}{\partial \varepsilon}=\frac{\partial \ln \mu}{\partial \ln  n}~,
\end{equation}
involves the second derivative of the pressure with respect to $\mu$ since $n=\partial P/\partial \mu$, 
see figure~\ref{fig:g2}.
The result is that for $k=1$ the function (\ref{mph_general}) exhibits a clear discontinuity in the speed of sound at $\varepsilon_{c}$ and $\varepsilon_{c}+\Delta\varepsilon$, whereas in between these borders, the speed of sound slightly increases relative to the case of the Maxwell construction for which $c_s^2=0$ in the mixed phase region. 
For $k=2$, the mixed phase pressure (\ref{mph_general}) allows for a continuous speed of sound. 
However, it is connected at $\varepsilon_{c}$ and $\varepsilon_{c}+\Delta\varepsilon$ to the speed of sound outside these borders with a jump in its derivative. 
At the order $k=3$ and higher the speed of sound behaves smoothly without a jump in its derivative.



\subsection{The mixing interpolation method (MIM)}
\label{MIM}
\begin{figure*}[!hbt]
\begin{center}$
\begin{array}{cc}
\includegraphics[width=0.5\textwidth]{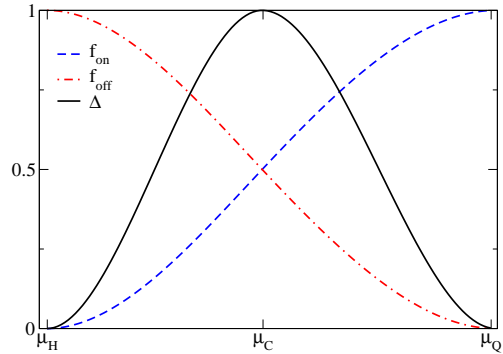}
\end{array}$
\end{center}
\caption{Polynomial switch functions $f_{\rm off/on}(\mu)$ as well as the function $\Delta(\mu)$.} 
  \label{fig:switch} 
\end{figure*}

This approach has recently been defined in Ref.~\cite{Alvarez-Castillo:2018rrv}, where the interpolation ansatz was based on trigonometric functions. 
Here we will use instead a polynomial ansatz for the interpolation that consists of a pair of functions $f_{\rm off} $ and $f_{\rm on} $ that will switch off and on the hadronic and quark parts of the equation of state, as well as an additional compensating function $\Delta$ in order to eliminate thermodynamic instabilities, see figure~\ref{fig:switch}. This interpolation is applied in the $p$ - $\mu$ plane within the range $\mu_{H} \leq \mu \leq \mu_{Q}$. 
%

The pressure that interpolates between the hadron and quark phase at the phase transition reads
\begin{equation}
P(\mu)=P_{H}(\mu)f_{\rm off}(\mu)+P_{Q}(\mu)f_{\rm on}(\mu)+\Delta(\mu) \Delta P.
\end{equation}

Even though  $f_{\rm off} $ and $f_{\rm on} $ might be any switching functions,  our choice of definition consists of the following pair of left and right side polynomials:
\begin{equation}
f_{>,L}=\alpha_{L}(\frac{\mu-\mu_{H}}{\mu_{Q}-\mu_{H}})^{2}+\beta_{L}(\frac{\mu-\mu_{H}}{\mu_{Q}-\mu_{H}})^{3}
\end{equation}
\begin{equation}
f_{<,R}=\alpha_{R}(\frac{\mu_{Q}-\mu}{\mu_{Q}-\mu_{H}})^{2}+\beta_{R}(\frac{\mu_{Q}-\mu}{\mu_{Q}-\mu_{H}})^{3}
\end{equation}
that together with the complementary functions $f_{>,R}(\mu)=1-f_{<,R}(\mu)$ and  $f_{<,L}(\mu)=1-f_{>,L}(\mu)$ will complete the switch functions.
The above coefficients  $\alpha_{L}$ , $\alpha_{R}$, $\beta_{L}$ and $\beta_{R}$ can be determined by the following conditions 
\[ 
\begin{array}{cccc}
   f_{{\lessgtr},L}(\mu)\Bigr|_{\mu=\mu_c} &=&f_{{\lessgtr},R}(\mu)\Bigr|_{\mu=\mu_c} &=1/2\\
      \frac{\partial f_{{\lessgtr},L}(\mu)}{\partial\mu}\Bigr|_{\mu=\mu_c} &=&\frac{\partial f_{{\lessgtr},R(\mu)}}{\partial\mu}\Bigr|_{\mu=\mu_c} & \\
       \frac{\partial^{2} f_{{\lessgtr},L}(\mu)}{\partial\mu^{2}}\Bigr|_{\mu=\mu_c} &=&\frac{\partial^{2} f_{{\lessgtr},R(\mu)}}{\partial\mu^{2}}\Bigr|_{\mu=\mu_c} &
\end{array} 
 \]
where the value of $1/2$ is chosen for symmetric convenience. 
Consequently, the switching functions are defined as
\[ f_{\rm on}(\mu) =  \left\{
\begin{array}{lr}
      0, &  \mu<\mu_{H}\\
   f_{>,L},& \mu_{H}\leq \mu\leq \mu_c \\
\end{array} 
\right. \]
\[ f_{\rm off}(\mu) =  \left\{
\begin{array}{lr}
f_{<,R},&   \mu_c\leq \mu\leq \mu_{Q} \\
0, &   \mu>\mu_{Q} \\
\end{array} 
\right. \]
and furthermore obey $f_{\rm off /on}(\mu)=1-f_{\rm on / off }(\mu)$.  

In order to construct a proper dimensionless function $\Delta(\mu)$ we introduce 
\[ \Delta(\mu) =  \left\{
\begin{array}{lr}
      0 & : \mu<\mu_{H}\\
   g_{L}(\mu) & : \mu_{H}\leq \mu\leq \mu_C \\
   g_{R}(\mu) & :  \mu_C\leq \mu\leq \mu_{Q} \\
      0 & :  \mu>\mu_{Q} \\
\end{array} 
\right. \]
consisting of the functions
\begin{equation}
g_{L}=\delta_{L}(\frac{\mu-\mu_{H}}{\mu_{C}-\mu_{H}})^{2}+\gamma_{L}(\frac{\mu-\mu_{H}}{\mu_{C}-\mu_{H}})^{3}
\end{equation}
\begin{equation}
g_{R}=\delta_{R}(\frac{\mu_{Q}-\mu}{\mu_{Q}-\mu_{C}})^{2}+\gamma_{R}(\frac{\mu_{Q}-\mu}{\mu_{Q}-\mu_{C}})^{3}
\end{equation}
 whose coefficients are determined by the conditions
 \[ 
\begin{array}{cccc}
   g_{L}(\mu)\Bigr|_{\mu=\mu_C} &=&g_{R}(\mu)\Bigr|_{\mu=\mu_C} &=1\\
    \frac{\partial g_{L}(\mu)}{\partial\mu}\Bigr|_{\mu=\mu_C} &=&\frac{\partial g_{R}(\mu)}{\partial\mu}\Bigr|_{\mu=\mu_C}&=0~.
\end{array} 
 \]
 
Regarding $\Delta P$ as the only free external parameter, up to this moment we have 10 unknowns and 8 independent equations which leave us with the possibility to fix the second order derivative of $P$ at $\mu_{H}$ and $\mu_{Q}$ in the following way: 
\[ 
\begin{array}{lr}
     \frac{\partial^{2}P}{\partial\mu^{2}}\Bigr|_{\mu=\mu_{H}} =&   \frac{\partial^{2}P_{H}}{\partial\mu^{2}}\Bigr|_{\mu=\mu_{H}} \\
      \frac{\partial^{2}P}{\partial\mu^{2}}\Bigr|_{\mu=\mu_{Q}}  =&   \frac{\partial^{2}P_{Q}}{\partial\mu^{2}}\Bigr|_{\mu=\mu_{Q}~. } 
\end{array} 
 \]

\section{Results}
\label{sec:results}
\subsection{Hybrid star EoS with mixed phases}

\begin{figure*}[!bhtp]
\begin{center}$
\begin{array}{cc}
\includegraphics[width=0.5\textwidth]{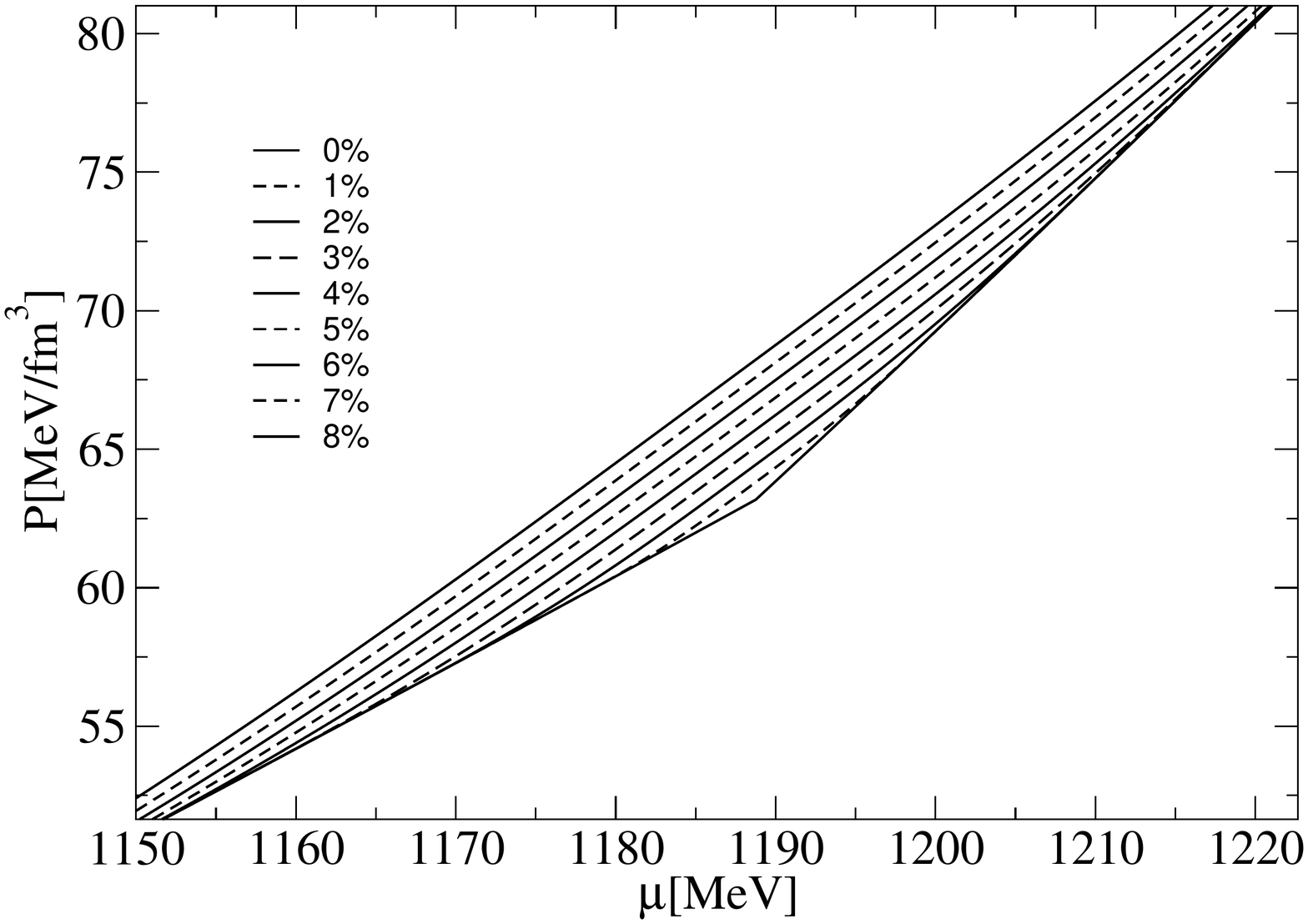} & \hspace{-0.5cm}\includegraphics[width=0.5\textwidth]{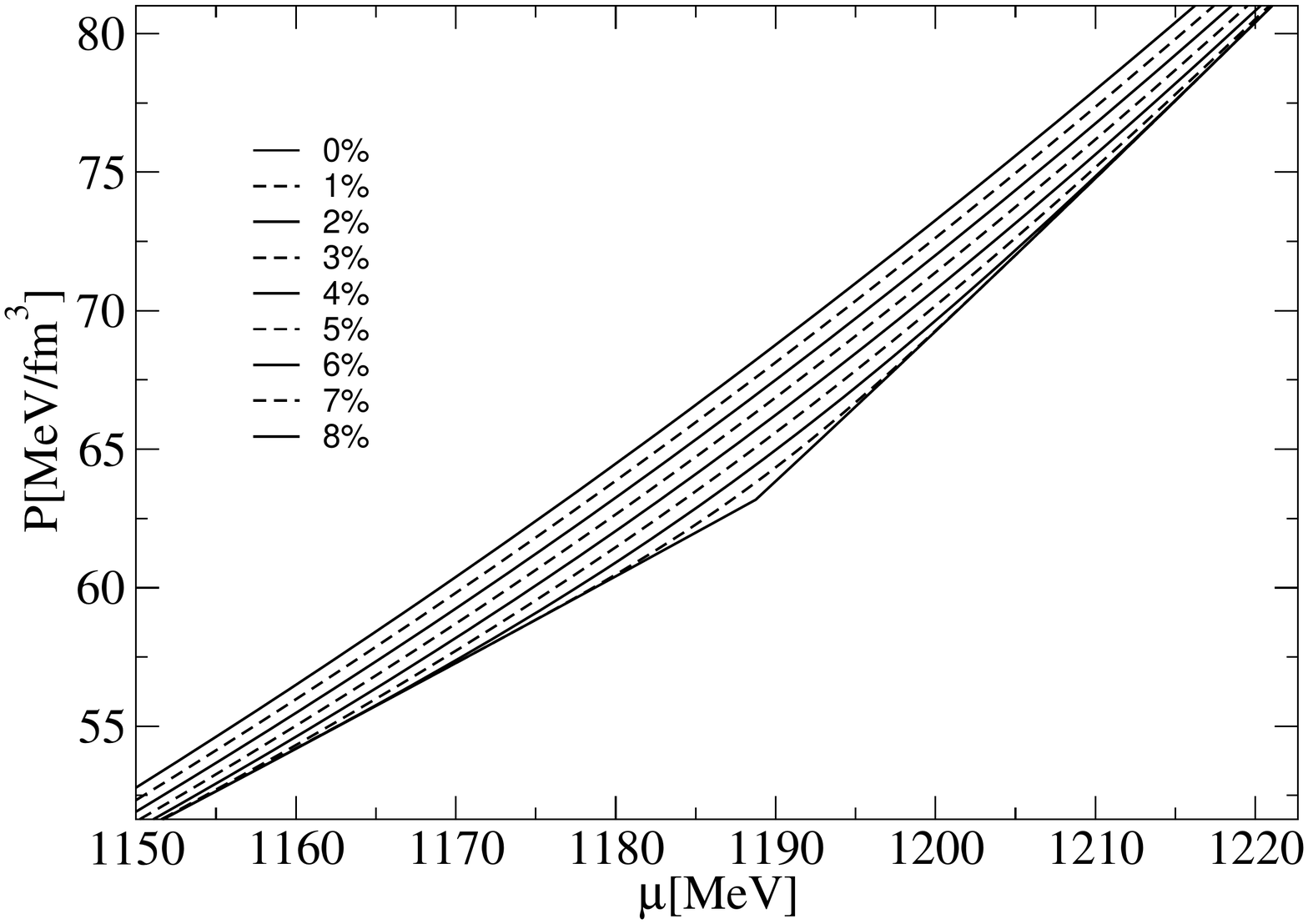}\\
\end{array}$
\end{center}
\caption{The EoS for pressure $P$ vs. chemical potential $\mu$ for both MIM (left panel) and RIM for a sixth order polynomial ansatz (right panel, $k=3$) approaches to the mixed phase construction. Different curves labelled by percentages correspond to values of 
	$\Delta_P=\Delta P/P_c$, where $\Delta_P=0$ corresponds to the Maxwell construction.} 
  \label{fig:PvsMu} 
\end{figure*}

\begin{figure*}[!bhtp]
\begin{center}$
\begin{array}{cc}
\includegraphics[width=0.55\textwidth]{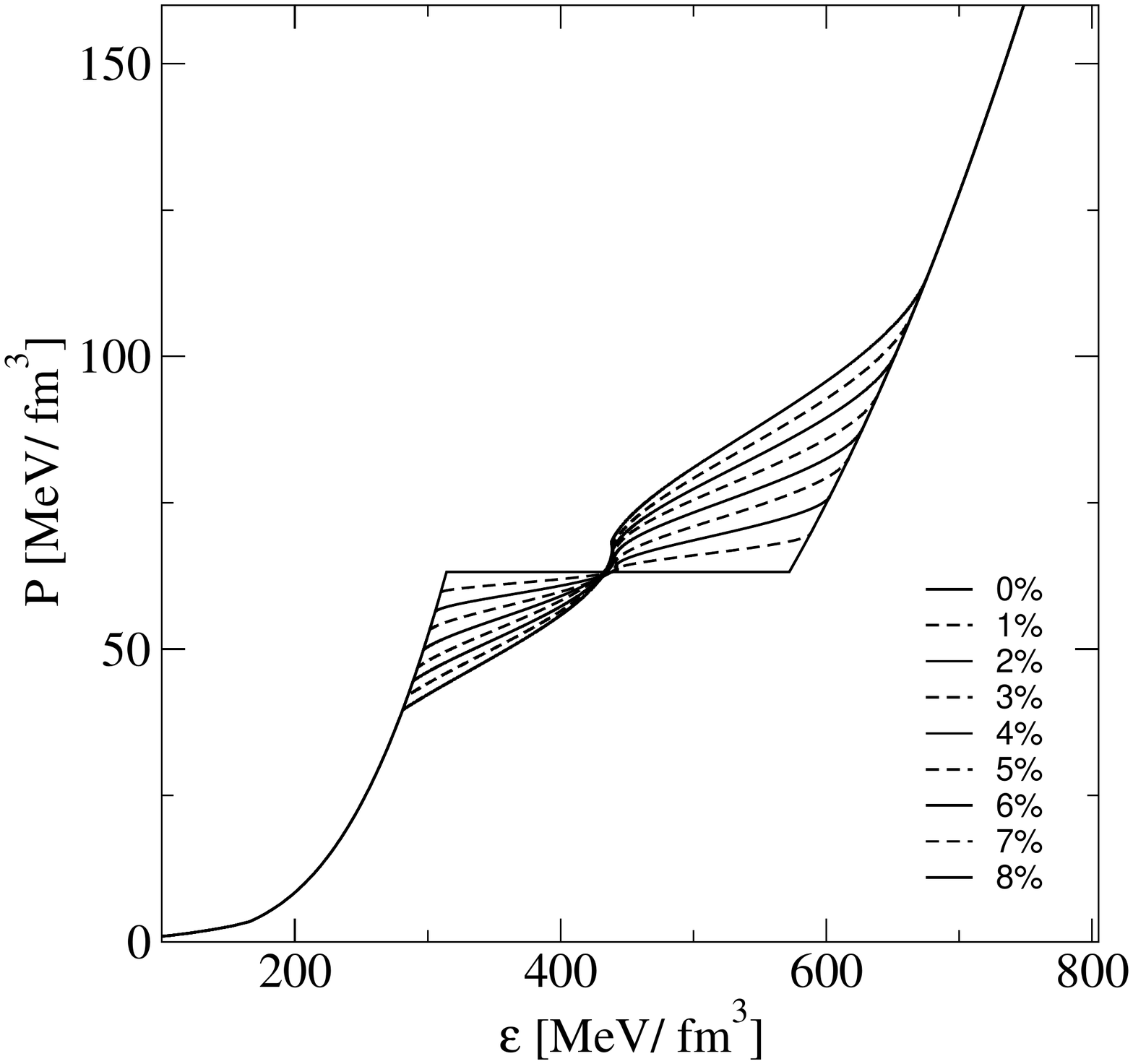} & \hspace{-1.0cm}\includegraphics[width=0.55\textwidth]{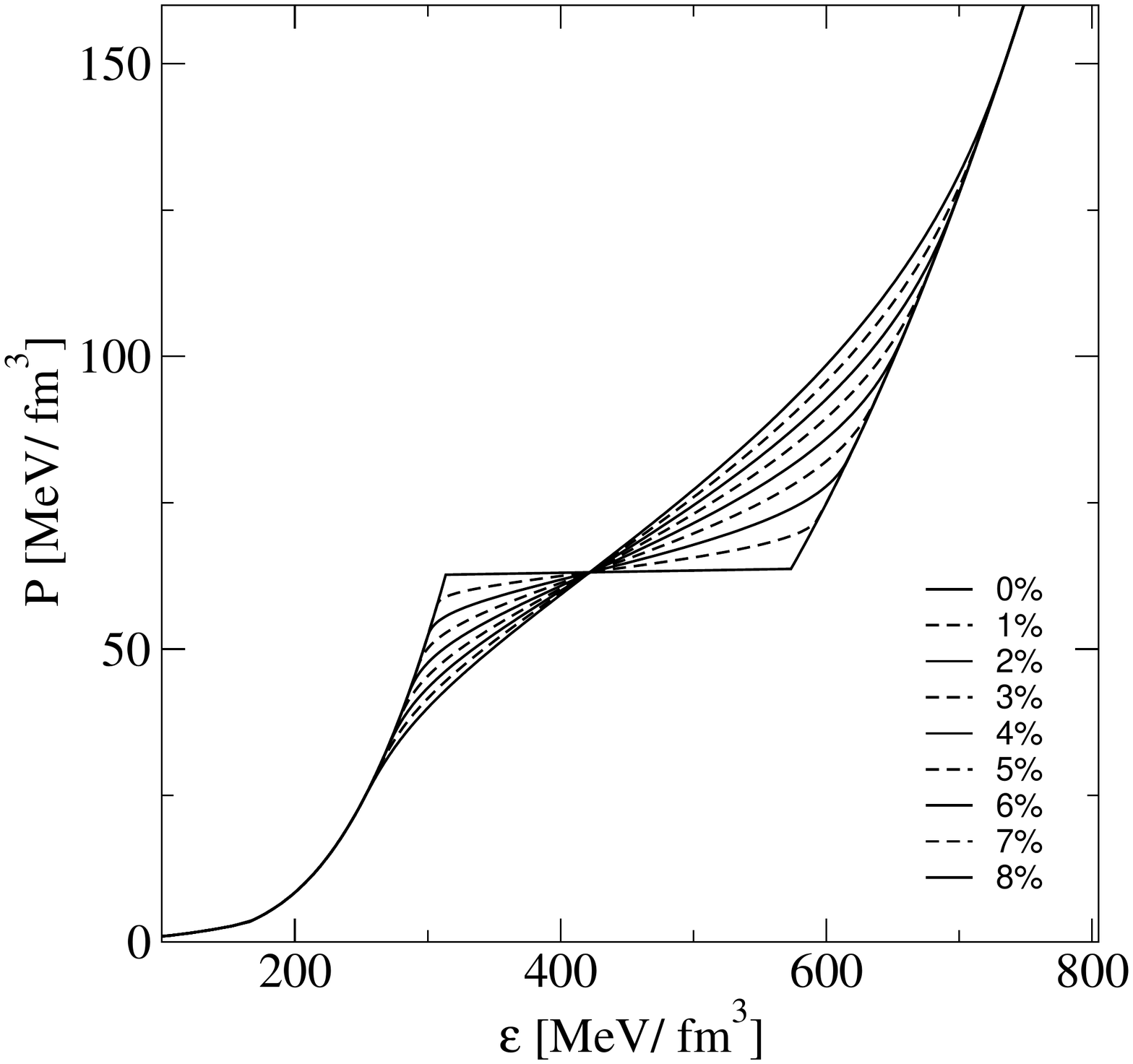}
\end{array}$
\end{center}
\caption{The EoS for pressure $P$ vs. energy density $\varepsilon$ for both MIM (left panel) and RIM for a sixth order polynomial ansatz (right panel, $k=3$) approaches to the mixed phase construction. Different curves labelled by percentages correspond to values of 
	$\Delta_P=\Delta P/P_c$, where $\Delta_P=0$ corresponds to the Maxwell construction.} 
  \label{fig:PvsEpsilon} 
\end{figure*}

\begin{figure*}[!bhtp]
\begin{center}$
\begin{array}{cc}
\includegraphics[width=0.5\textwidth]{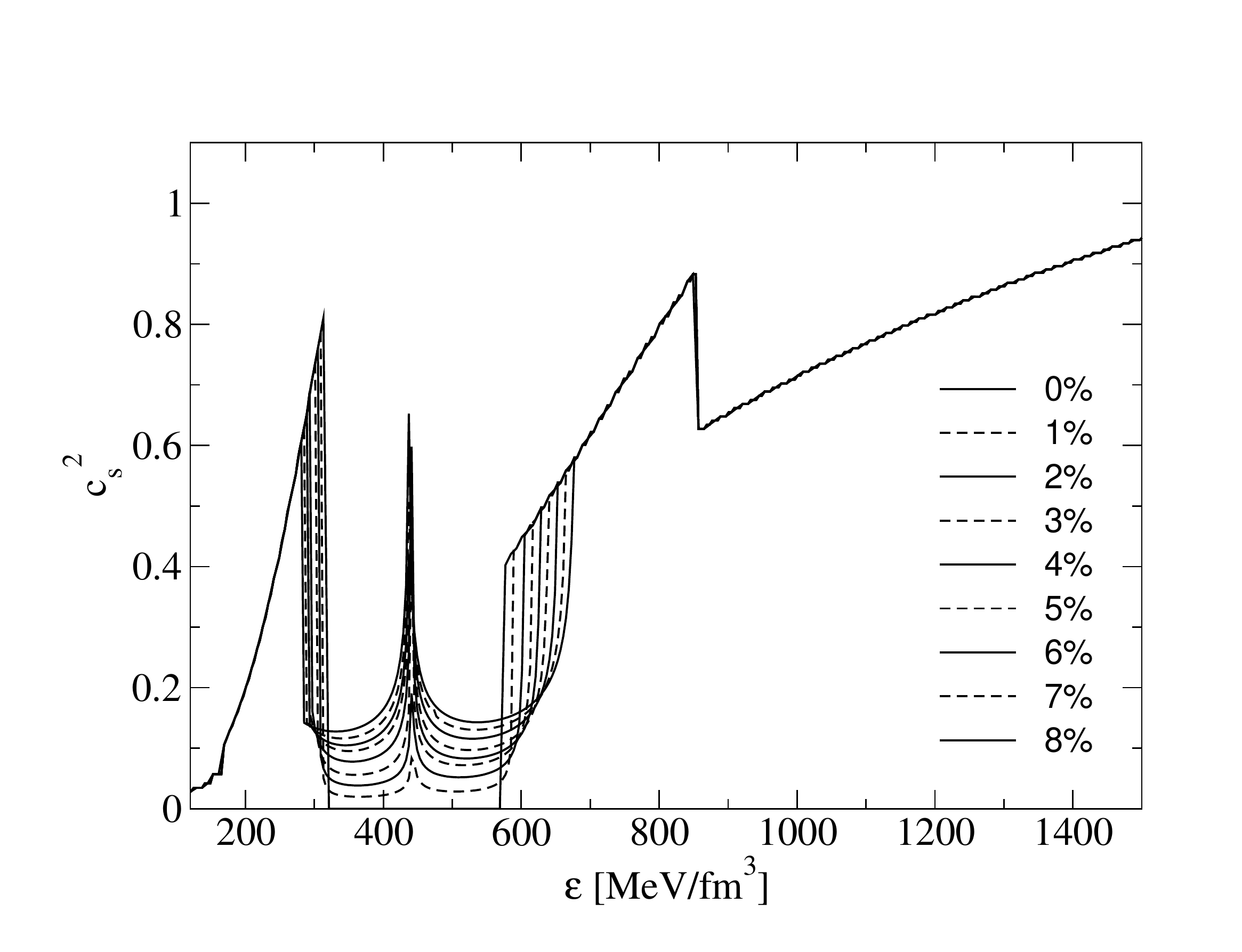} & \hspace{-0.5cm}\includegraphics[width=0.5\textwidth]{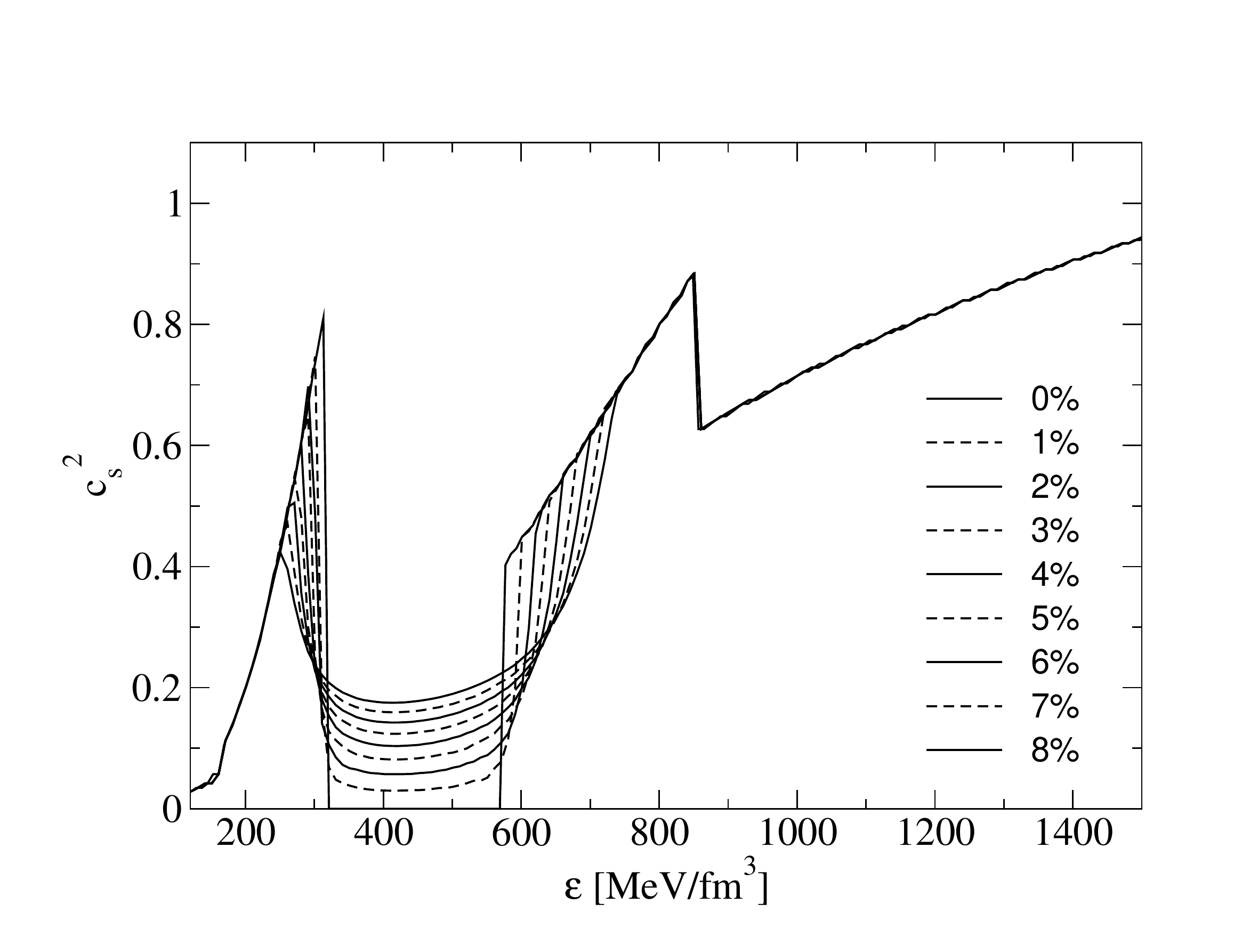}\\
\end{array}$
\end{center}
\caption{The squared speed of sound $c_s^2$ vs. energy density $\varepsilon$ for both MIM (left panel) and RIM (right panel, $k=3$) approaches to the mixed phase construction. 
Different curves labelled by percentages correspond to values of $\Delta_P=\Delta P/P_c$, where $\Delta_P=0$ corresponds to the Maxwell construction. A clear feature of the MIM that distinguishes it from the RIM is the intermediate stiffening of the EoS, apparent by the peaked structure inside the mixed phase region. See~\cite{Alvarez-Castillo:2018rrv} for discussion.} 
  \label{fig:SoS} 
\end{figure*}

The two interpolation methods presented above result in a thermodynamically consistent EoS. 
Knowing that $n=\partial P/\partial \mu$, the thermodynamic identity used to derived all the needed variables at zero temperature reads
\begin{equation}
 \varepsilon=-p + \mu \, n.
\end{equation}
Figure~\ref{fig:PvsMu} shows the resulting mixed phase interpolations for both approaches charaterised by the dimensionless pressure increment $\Delta_P=\Delta P/P_c$ that ranges from $1\%$ to $8\%$, where $\Delta_P=0$ reproduces the Maxwell construction. 
Figure~\ref{fig:PvsEpsilon} shows pressure values depending on energy density. The first order phase transition via a Maxwell construction corresponds to the $\Delta P=0$ case with the pressure being constant in the mixed phase region. 
Furthermore, figure~\ref{fig:SoS} shows the squared speed of sound for both approaches where the difference between them becomes obvious: while the MIM shows a peak in the mixed phase region the RIM shows a rather structureless behaviour in this region. 
This feature is a direct consequence of the functional form of the interpolation implemented by the two methods.

\subsection{Compact star sequences}

\begin{figure*}[!bhtp]
\begin{center}$
\begin{array}{cc}
\includegraphics[width=0.65\textwidth]{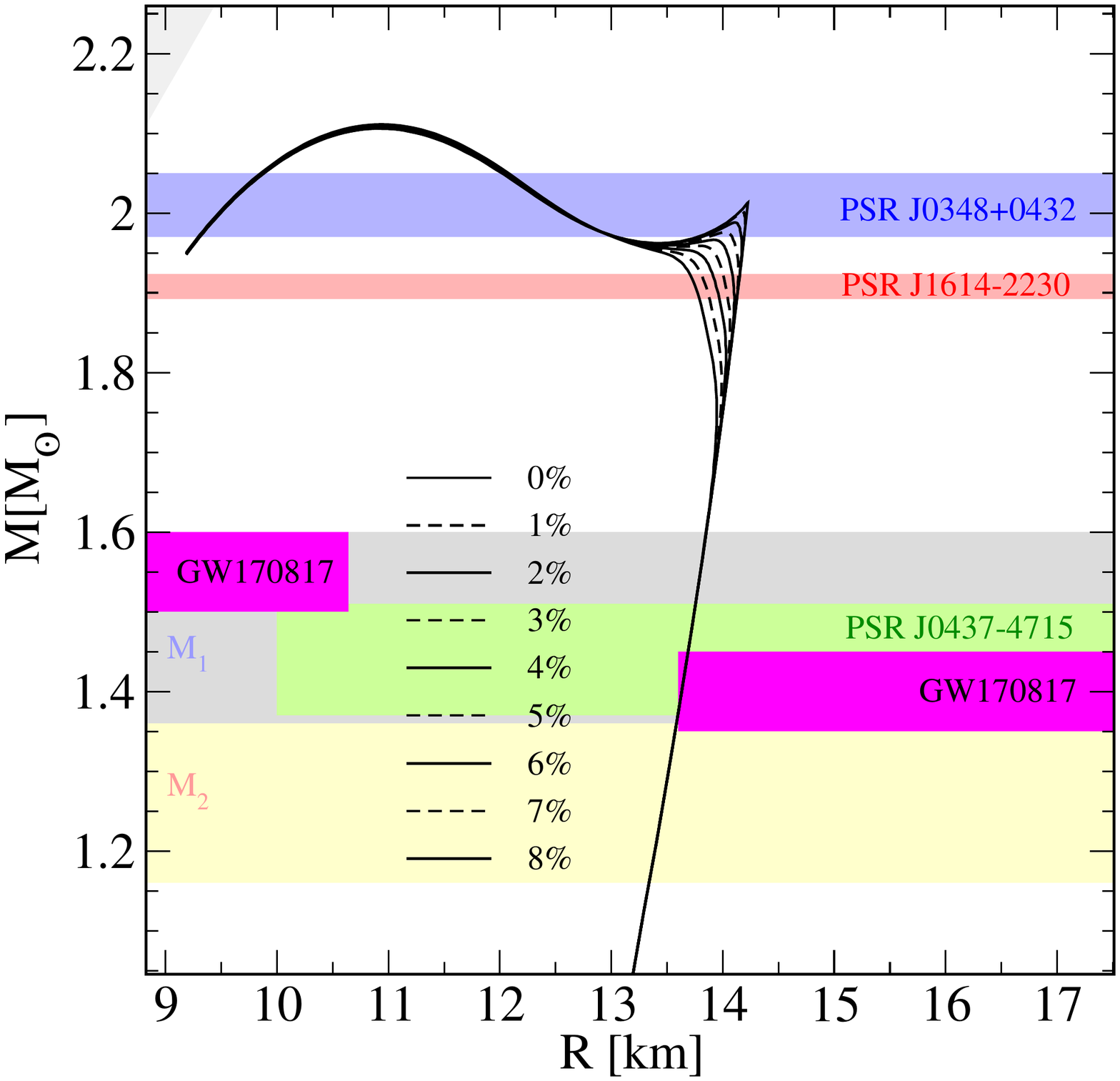} &\hspace{-3cm} \includegraphics[width=0.65\textwidth]{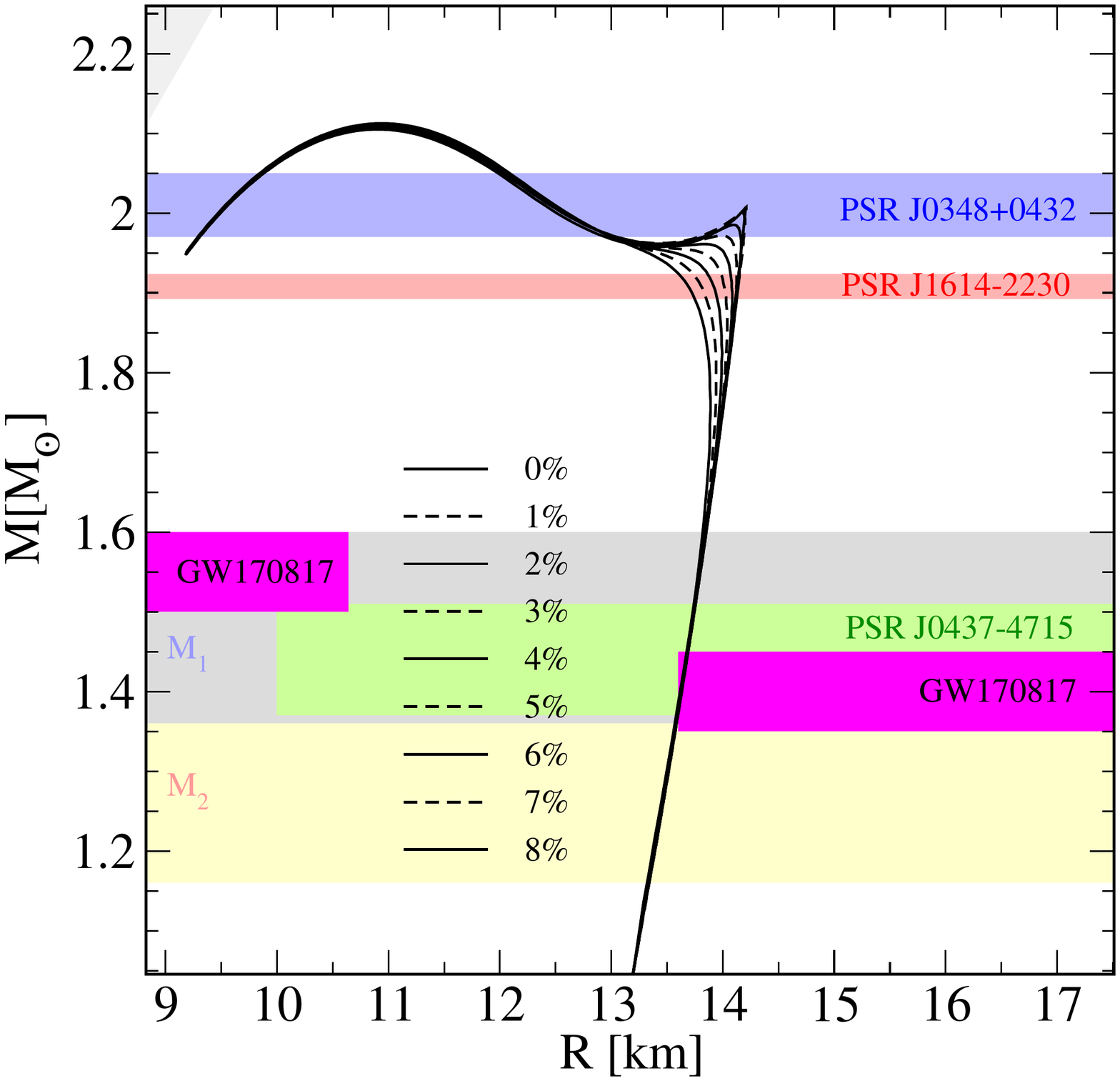}
\end{array}$
\end{center}
\caption{\label{MvsR}
Mass-radius relations for both mixed phase approaches, MIM (left panel) and RIM for a sixth order polynomial ansatz (right panel). 
Each curve corresponds to an EoS with a chosen $\Delta P$ value given as a percentage of the critical Maxwell pressure $P_c$ represented by alternating line-styles.
The shaded areas correspond to compact star measurements: 
The blue and red horizontal bands correspond to mass measurements of  PSR J1614-2230~\cite{Antoniadis:2013pzd} and PSR J0348+432~\cite{Arzoumanian:2017puf}, respectively. 
The gray and orange bands denoted by M1 and M2 are the compact star mass windows for the binary merger GW170817.
The green band corresponds to the $1.44\pm0.07$ M$_{\odot}$ mass of PSR J0437-4715 whose radius is expected to be measured by NICER~\cite{Arzoumanian:2009qn}.
The hatched regions are excluded by GW170817:
the star radius at 1.6 M$_{\odot}$ cannot be smaller than 10.68 km~\cite{Bauswein:2017vtn} and for a 1.4 M$_{\odot}$ the star has to have a radius smaller than 13.6km~\cite{Annala:2017llu}. 
The maximum mass of compact stars shall be below $2.16$~M$_\odot$ \cite{Rezzolla:2017aly}.
} 
\end{figure*}

\begin{figure*}[!bhtp]
\begin{center}$
\begin{array}{cc}
\includegraphics[width=0.5\textwidth]{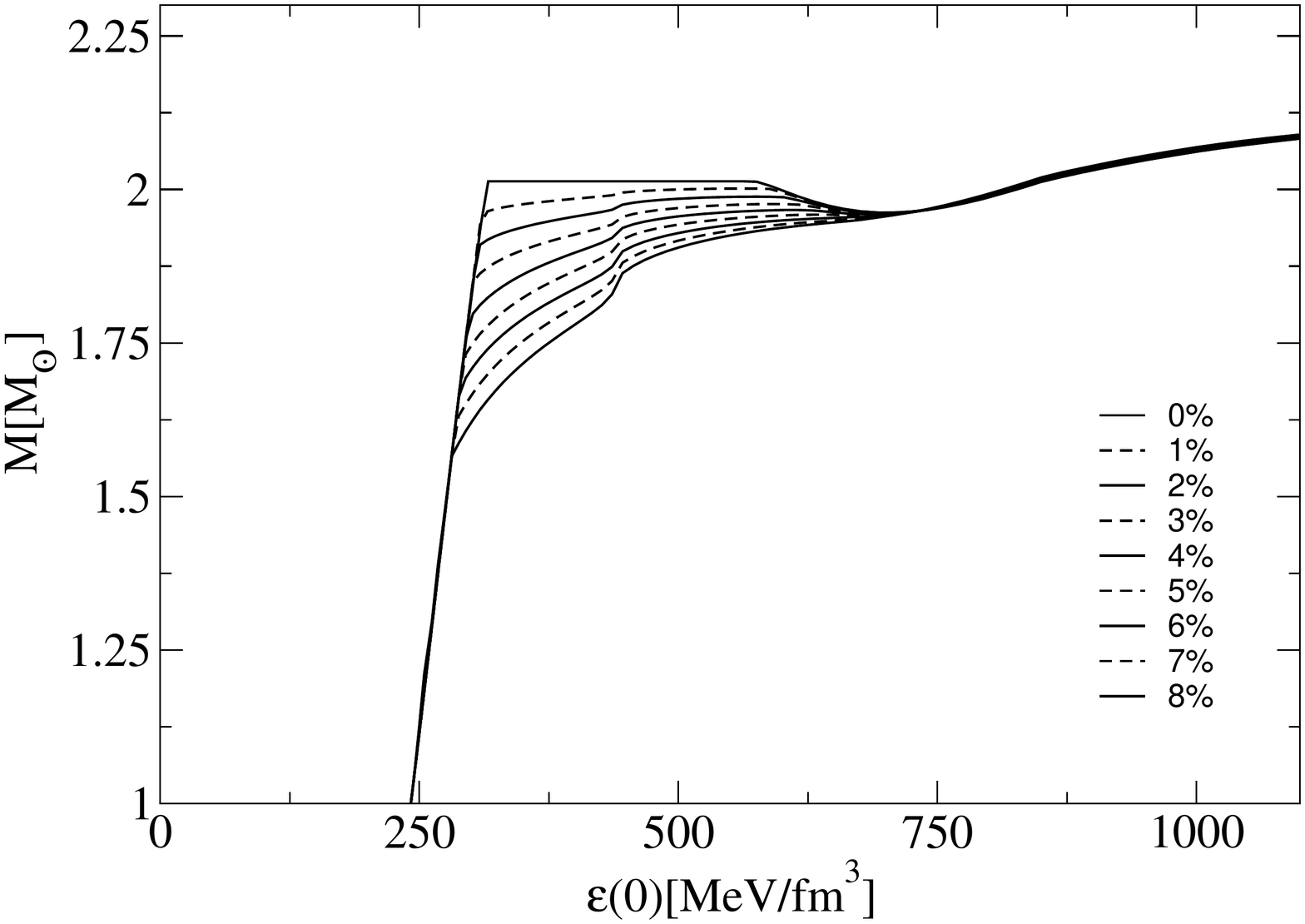} & \hspace{-1cm}\includegraphics[width=0.5\textwidth]{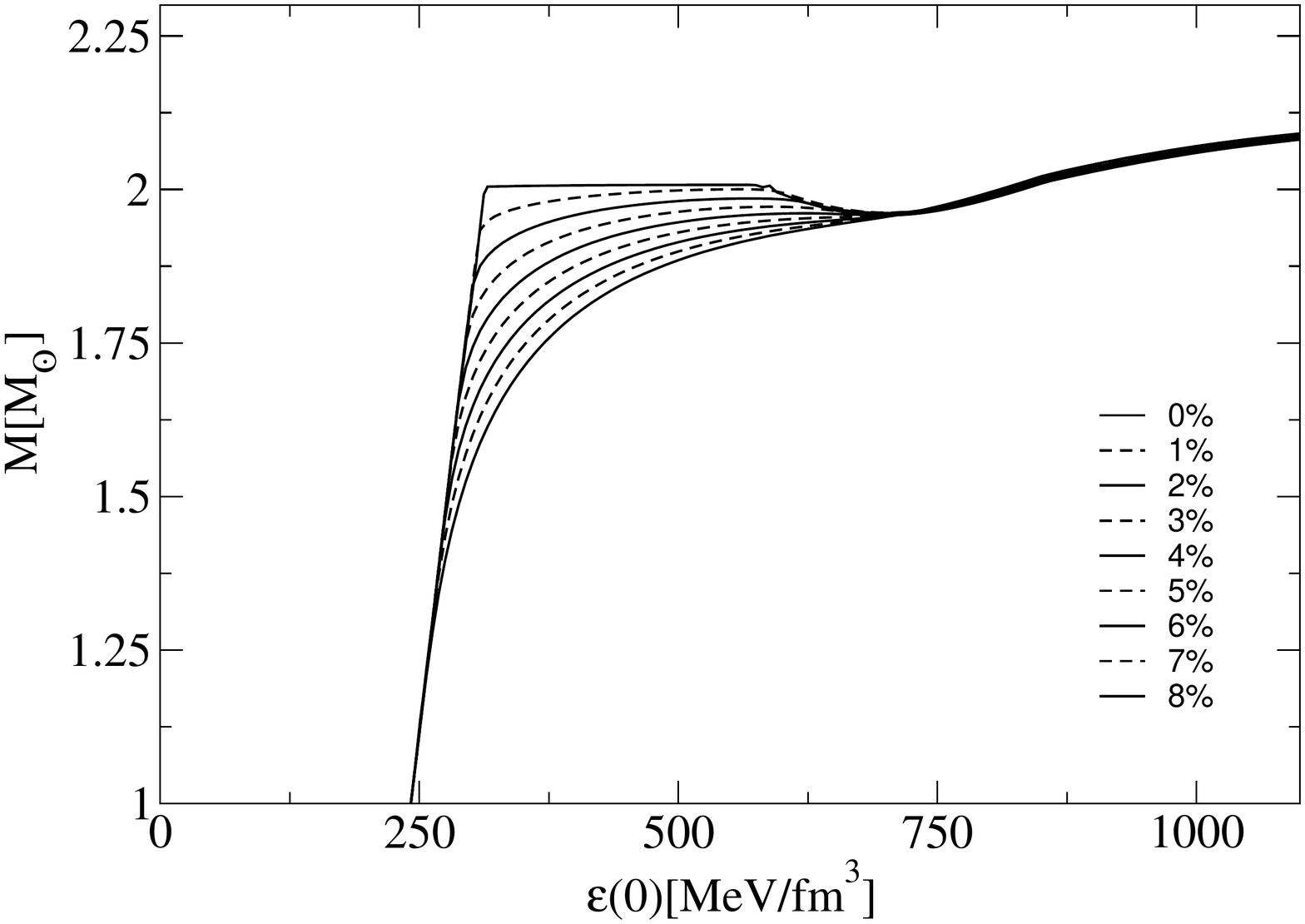}\\
\includegraphics[width=0.5\textwidth]{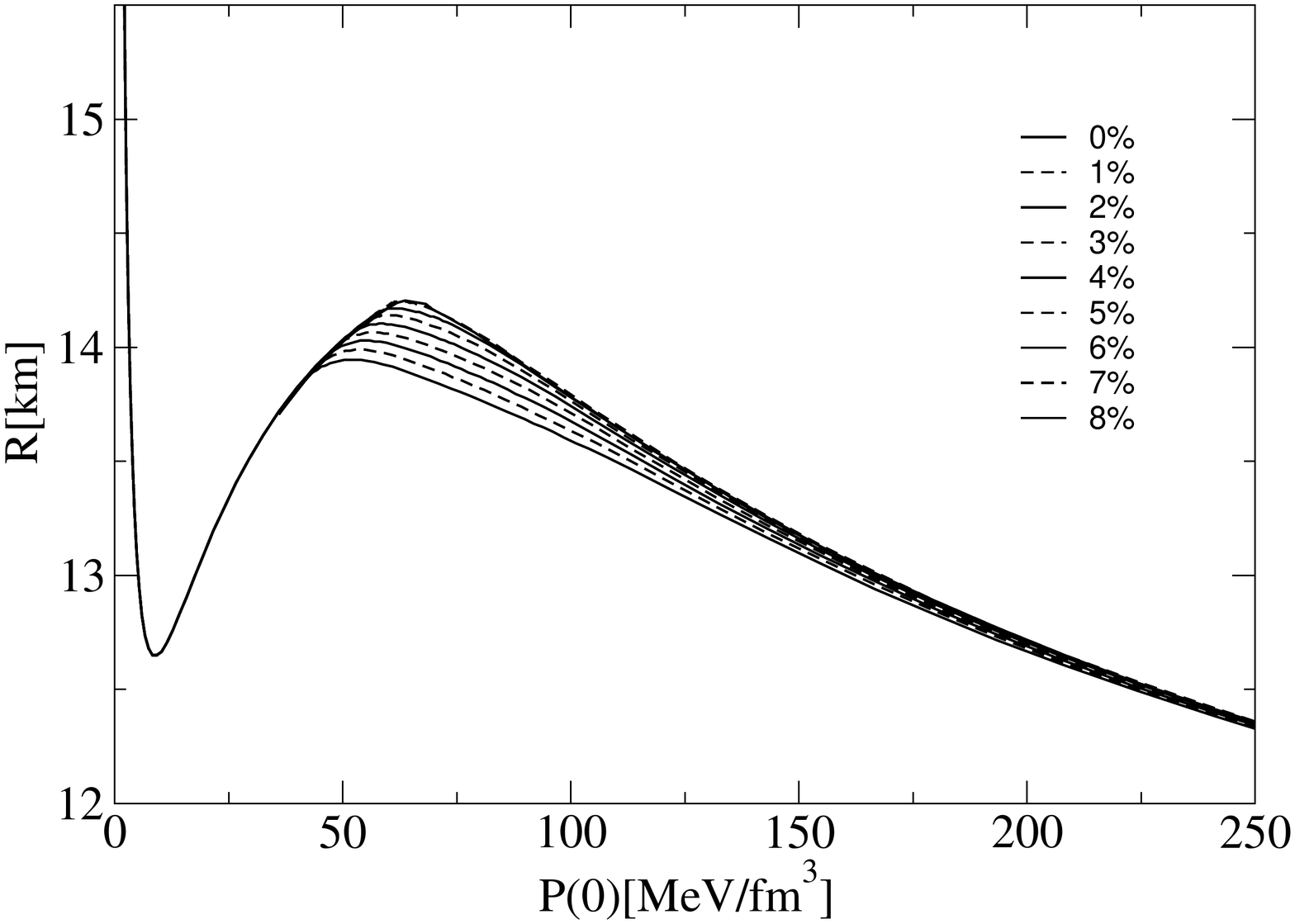} & \hspace{-1cm}\includegraphics[width=0.5\textwidth]{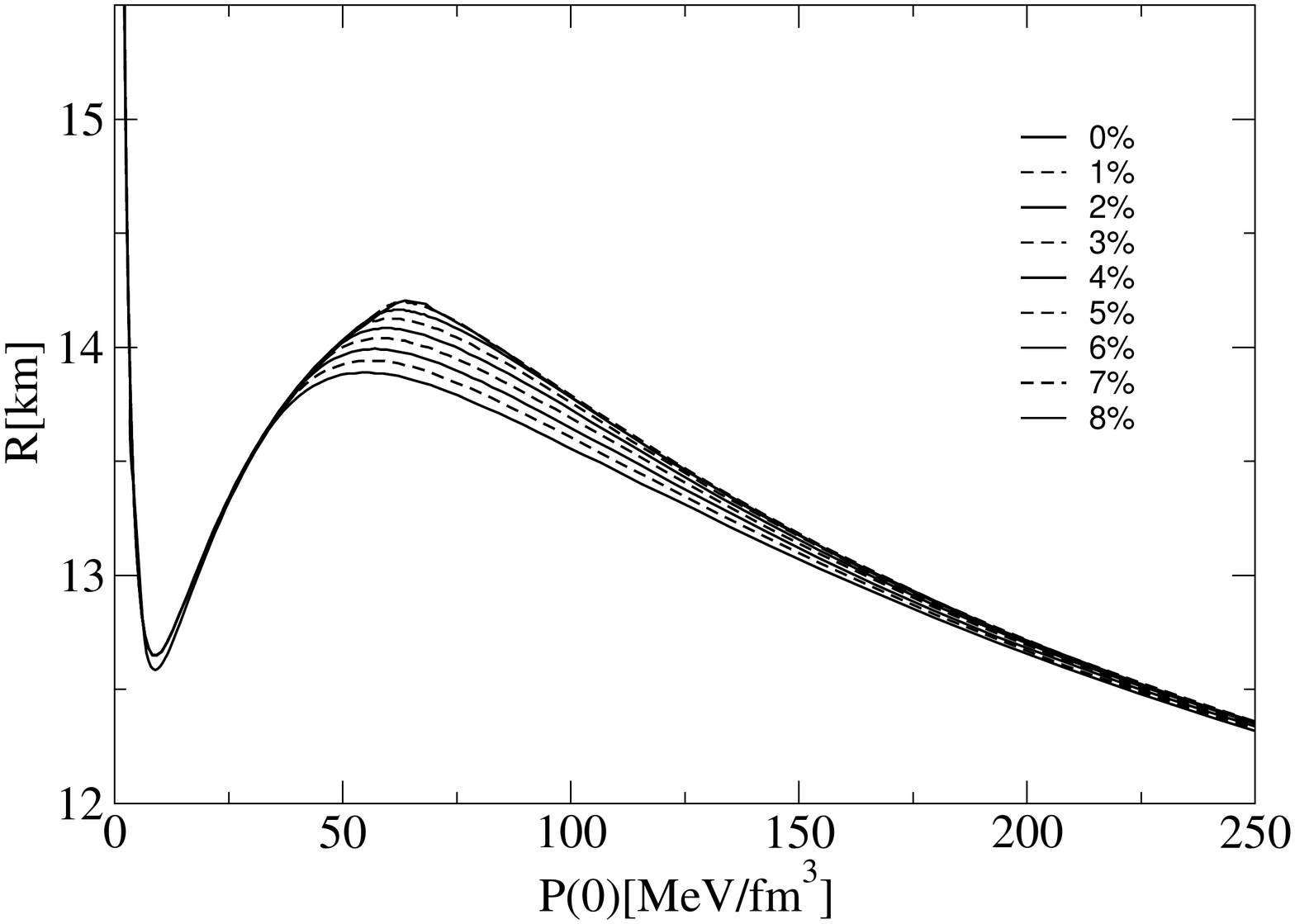}
\end{array}$
\end{center}
\caption{\label{fig:MRvsnc}
\textit{Upper panel}: Mass as a function of central energy density for both mixed phase approaches, MIM (left panel) and RIM (right panel). 
\textit{Lower panel}: Radius as a function of central pressure for all MIM (left panel) and RIM (right panel, $k=3$) sequences.
Each curve corresponds to an EoS with a chosen $\Delta P$ value given as a percentage of the critical Maxwell pressure $P_c$ represented by alternating line-styles. The case $\Delta P=0$ corresponds to the Maxwell construction which produces a sharp edge in the curves.} 
\end{figure*}

In order to compute the compact star internal pressure (energy density) profiles leading to mass-radius relations, we solve the  Tolman--Oppenheimer--Volkoff (TOV) equations~\cite{Tolman:1939jz,Oppenheimer:1939ne} derived in the framework of General Relativity for a static, spherically-symmetric compact star
\bea
\frac{dP( r)}{dr}&=& 
-\frac{G\left(\varepsilon( r)+P( r)\right)
\left(M( r)+ 4\pi r^3 P( r)\right)}{r\left(r- 2GM( r)\right)},\\
\frac{dM( r)}{dr}&=& 4\pi r^2 \varepsilon( r),
\eea
with the boundary conditions $P(r=R)=0$, M(0)=0 and $M(R)=M$ that serve to determine the total stellar mass $M$  and total stellar radius $R$  once a central pressure $ P(0)= P(r=0)$ (and with it the central energy density because $P(\varepsilon)$ is known) is given as input. 
By increasing the central energy density values, a whole sequence of star configurations up to the one with the maximal mass can be obtained, thus populating the mass-radius diagram. 
Figure~\ref{MvsR} shows compact star sequences for all models characterised by the $\Delta P$ value for both, the MIM and RIM approaches together with up-to-date constraints from astrophysical measurements. 
We can notice that for the lower values of $\Delta P < 6\%$  the HMT phenomenon persists regardless which mixed phase interpolation method has been applied.
In Fig.~\ref{fig:MRvsnc} we show the mass versus central energy density and the radius versus central pressure for both interpolation methods.
For the MIM one observes a trace of the intermediate stiffening effect in the mass versus central energy density which is absent for the RIM.

\begin{figure*}[!bhtp]
\begin{center}$
\begin{array}{cc}
\includegraphics[width=0.5\textwidth]{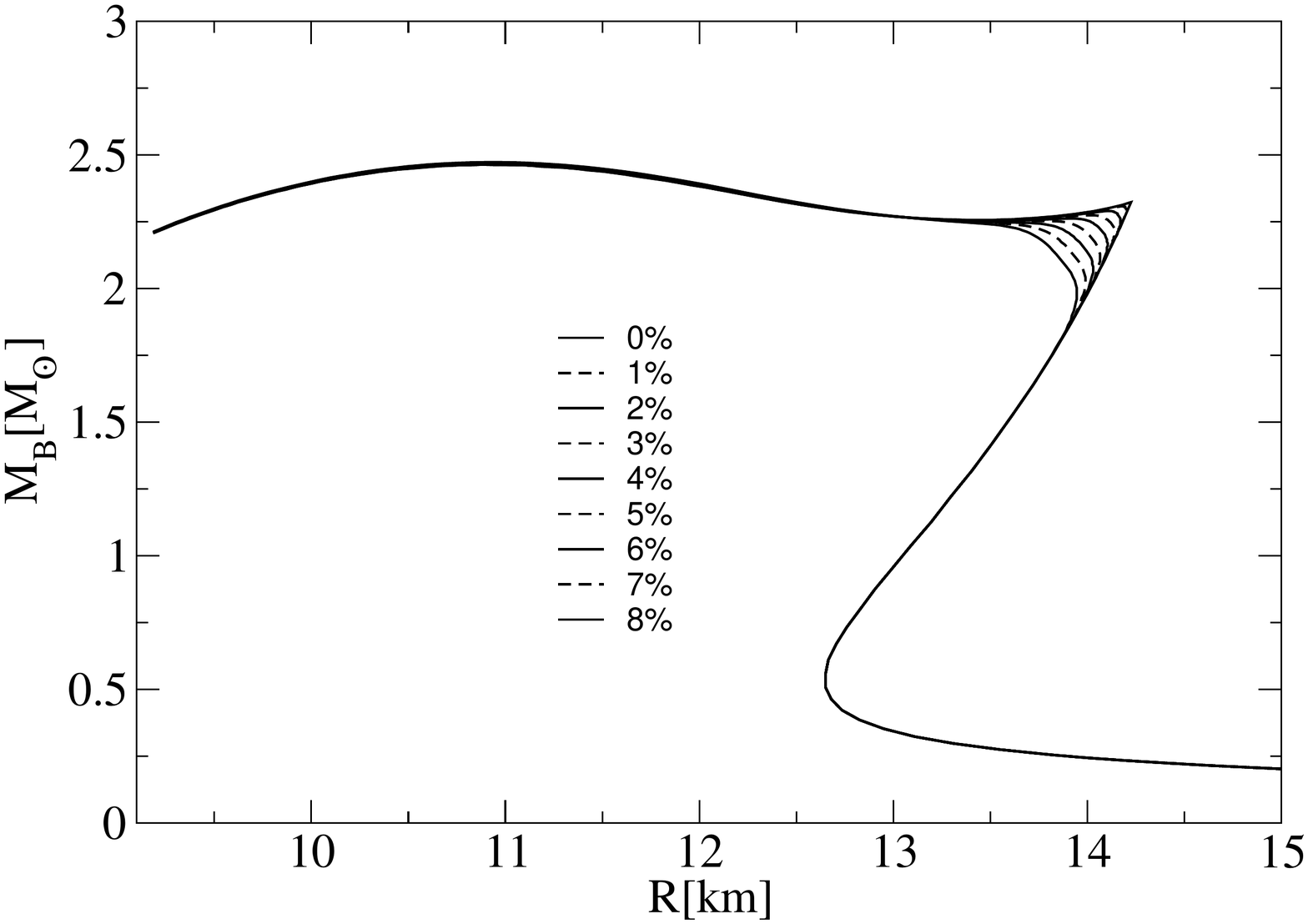} & \hspace{-1cm}\includegraphics[width=0.5\textwidth]{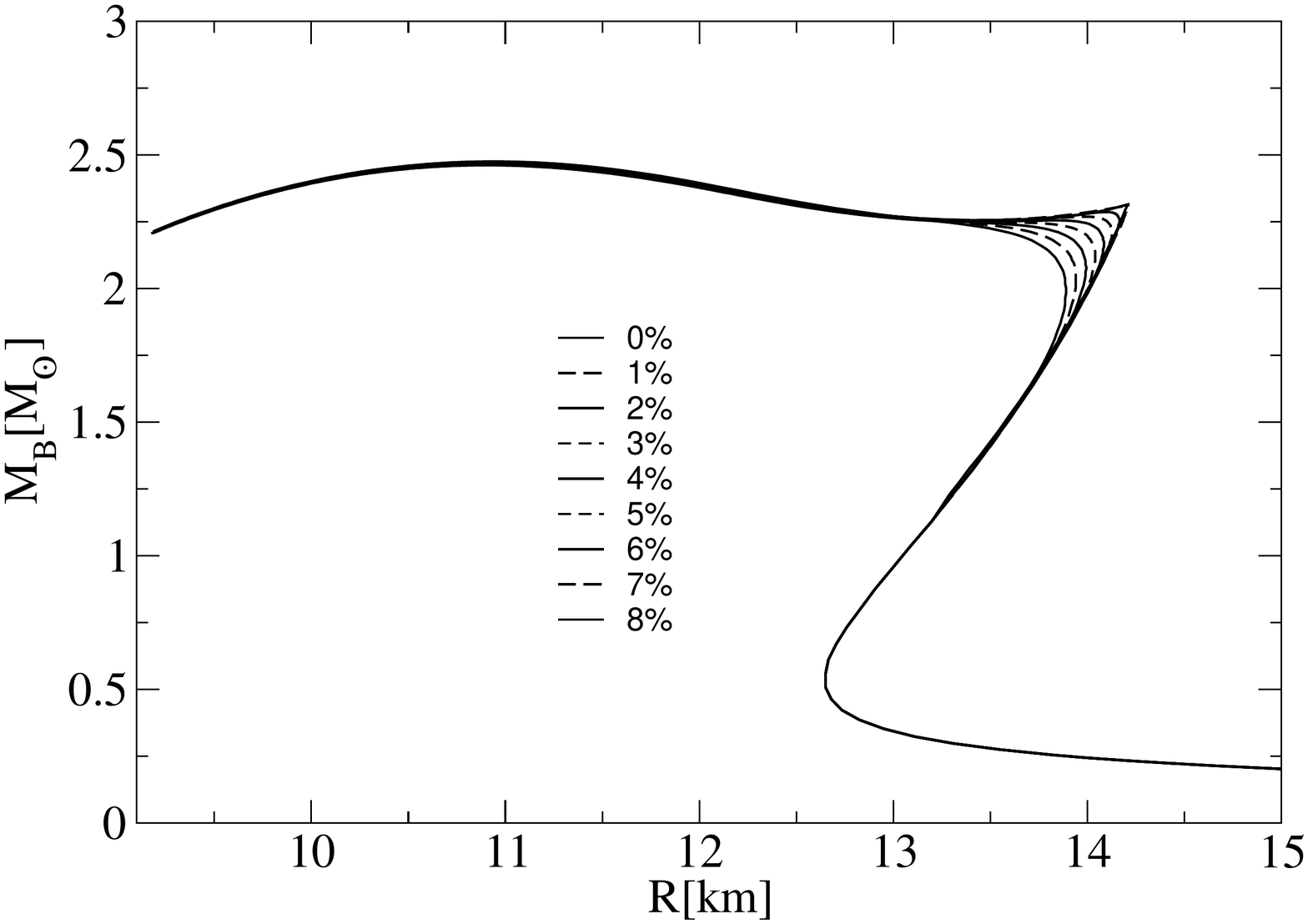}\\
\includegraphics[width=0.5\textwidth]{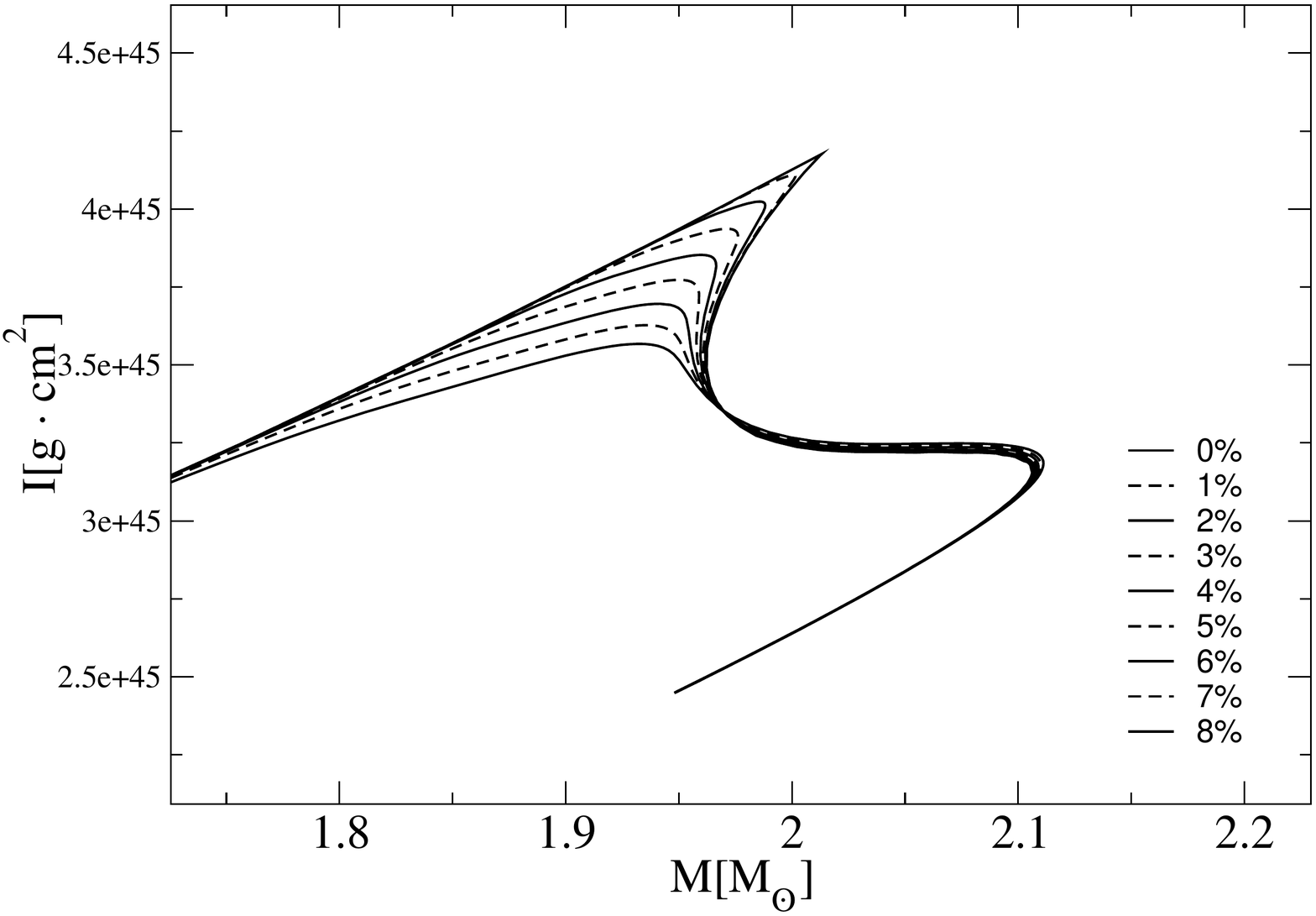} & \hspace{-1cm}\includegraphics[width=0.5\textwidth]{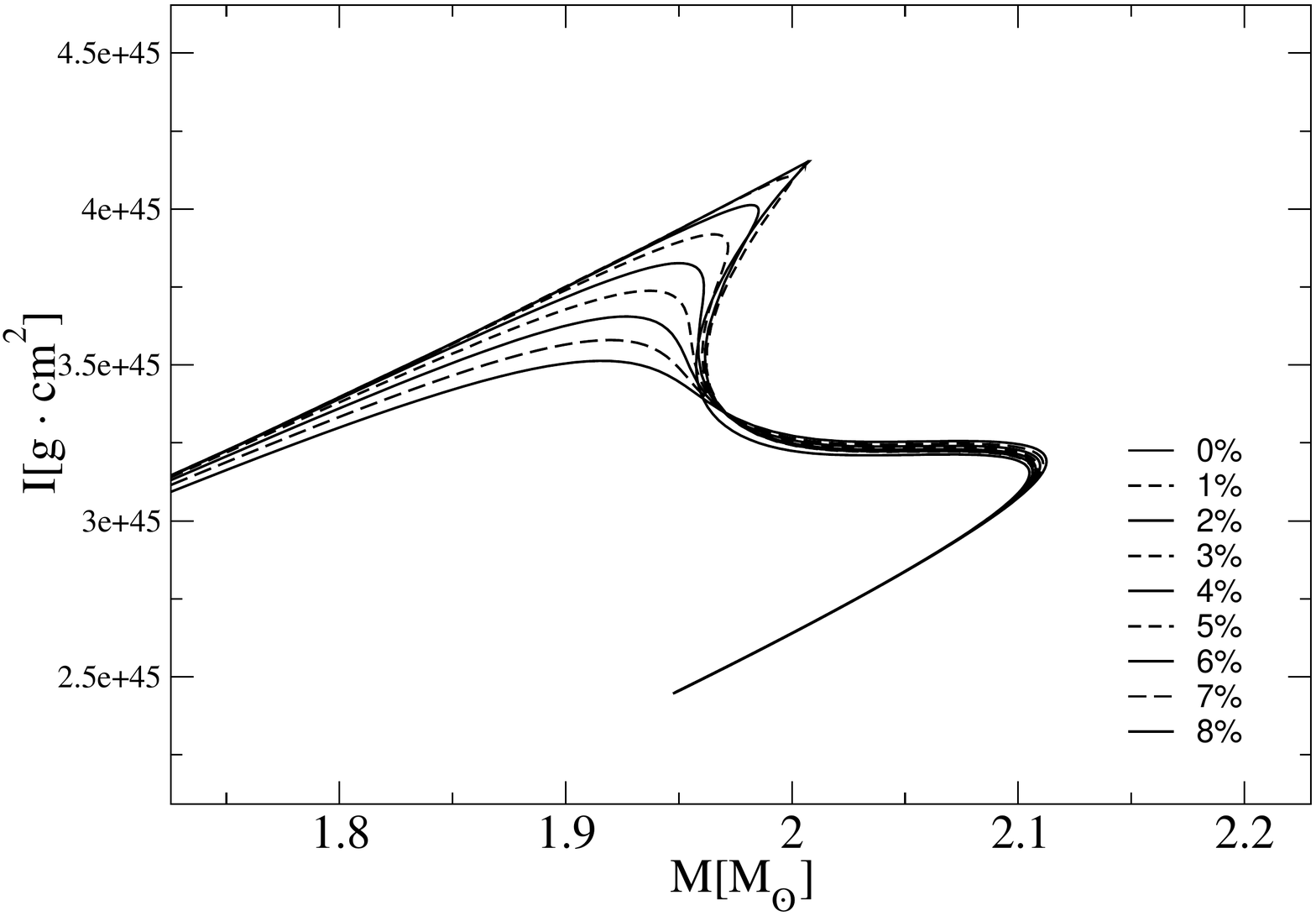}
\end{array}$
\end{center}
\caption{\label{MbvsR}
\textit{Upper panel}: Baryonic Mass vs. radius for both mixed phase approaches, MIM (left panel) and RIM (right panel, $k=3$). \textit{Lower panel}: Moment of inertia as a function of total mass for MIM (left panel) and RIM (right panel, $k=3$) approaches.
Each curve corresponds to an EoS with a chosen $\Delta P$ value given as a percentage of the critical Maxwell pressure $P_c$ represented by alternating line-style values. The case $\Delta P=0$ corresponds to the Maxwell construction which produces a sharp edge in the curves.} 
\end{figure*}

In addition, two other quantities of astrophysical interest are the total baryonic mass of the star that results from integrating the following equation 
\bea
\frac{d N_B( r)}{dr}&=& 4\pi r^2 \left(1-\frac{2GM( r)}{r}\right)^{-1/2}n( r),
\eea
and similarly, its moment of intertia~\cite{Ravenhall:1994}
\begin{equation}
 I \simeq \frac{J}{1+2GJ/R^{3}c^{2}} ,~~J=\frac{8\pi}{3}\int_{0}^{R}r^{4}\left(\rho+\frac{p}{c^2}\right)\Lambda dr,~~\Lambda=\frac{1}{1-2Gm/rc^{2}},
\end{equation}
which are related to observational phenomena as well, like energetic emissions that might conserve baryon mass or moment of inertia dependent pulsar glitches. 
For a detailed discussion of the moment of inertia in the
slow-rotation approximation, and for the hybrid star case
see, e.g., \cite{Chubarian:1999yn,Zdunik:2005kh,Bejger:2016emu}, and references therein.
In figure~\ref{MbvsR} we show the baryon mass versus radius and and the moment of inertia versus gravitational mass for the compact star sequences obtained in this work with both interpolation methods.
When increasing the pressure increment from $\Delta_P=0$ to $8\%$, the sharp edges which are obtained for the Maxwell construction case get washed out.
One observes no qualitative difference between the MIM and the RIM in the patterns of these families of sequences.
For $\Delta_P>5\%$, the second and third family branches in the $M_B$ vs. $R$ diagrams get joined so that neutron star and hybrid star configurations form a connected sequence and the HMT phenomenon get lost. 
This effect is reflected in the $I$ vs. $M$ diagrams by the loss of multi-valuedness (the lowest branch up to the maximum mass of $2.11$~M$_\odot$ shall be ignored because it is unstable). 
From the $M_B$ vs. $R$ diagrams one can read off which configuration on the hybrid star branch would be reached when the maximum mass neutron star configuration would collapse under conservation of baryon number. Comparing the gravitational masses of these two star configurations one can estimate the release of binding energy in this process, see Ref.~\cite{Bejger:2016emu}.


\section{Conclusions}
\label{sec:conclude}

In this work we have introduced two interpolation approaches to a mixed phase at the hadron-quark phase transition. 
An advantage of these two interpolation methods presented here over the construction employing hyperbolic tanges functions~\cite{Alvarez-Castillo:2014dva,Alvarez-Castillo:2017xvu} is the finite extension in chemical potentials of the mixed phase between the hadronic and the quark EoS, whereas the latter strictly converges only at infinity.

While each approach uses a different functional form both of them fulfill the same conditions at the border of the mixed phase. 
We have found that both methods can be distinguished by the behaviour of the speed of sound that they predict.  
The MIM approach motivated by the analogy with sequential phase transitions occurring for substitutional compounds in the neutron star crust finds an intermediate stiffening of the mixed phase EoS. The RIM approach does not exhibit this feature.
In the case of the RIM approach, we have studied both, a fourth and sixth order polynomial interpolation.
We found that the latter connects the hadron and quark EoS smoothly up to second derivatives which is being visible in smooth behaviour of the speed of sound. 
However, the differences in the neutron star properties for both polynomial orders are safely negligible.

The macroscopic properties of compact stars show for both mixed phase constructions a very similar systematic behaviour as the pressure increment $\Delta P$ is increased: the mass-radius relation smoothens out eliminating the gap between second and third branches, but only for the highest values we have considered. 
Up to $\Delta_P\sim 5\%$ the HMT phenomenon is robust against the mixed phase construction, regardless whether the MIM or RIM approach is used.
For the mass versus central energy density, one observes a trace of the intermediate stiffening effect for the MIM which is absent for the RIM.
For the other compact star quantities evaluated here, the baryonic mass and the moment of inertia, both interpolation methods display a similar type of behaviour when the pressure increment is varied.

The methods presented here can potentially be applied to the compact star crust-core transition as well. Just like at the hadron-quark boundary, the transition at the bottom of the crust may proceed via pasta phases dominated by Coulomb forces and surface tension effects \cite{Ravenhall:1983uh}. 
Further astrophysical aspects of mixed phases inside neutron stars include potentially observable effects such as the rotational evolution, pulsar glitches, gravitational wave emission and cooling. 
They could be sufficiently sensible to the nature of the phase transition, proceeding via pasta phases or not, and thus provide potential signatures of the presence and extension of a mixed phase in compact stars.

\acknowledgments{
D.A.-C. acknowledges support from the Bogoliubov-Infeld program for collaboration of JINR Dubna with Polish Academic Institutions.
The research was carried out with financial support by the Russian Science Foundation under project \#17-12-01427)

}

\reftitle{References}




\end{document}